\documentclass[lettersize,journal]{IEEEtran}
\usepackage{amsmath,amsfonts}
\usepackage{algorithmic}
\usepackage{algorithm}
\usepackage{array}
\usepackage[caption=false,font=normalsize,labelfont=sf,textfont=sf]{subfig}
\usepackage{textcomp}
\usepackage{stfloats}
\usepackage{tabularray}
\usepackage{url}
\usepackage{verbatim}
\usepackage{graphicx}
\usepackage{cite}
\usepackage{subfig}
\usepackage{hyperref}
\usepackage[normalem]{ulem}
\usepackage{tabularray}
\usepackage{booktabs}
\usepackage{multirow}
\usepackage{tabularx}
\usepackage{array}
\usepackage{makecell}
\usepackage[normalem]{ulem}

\newcolumntype{Y}{>{\centering\arraybackslash}X}
\newcolumntype{C}[1]{>{\centering\arraybackslash}p{#1}}

\begin{document}

\title{M3F-UAV: A Missing-Modality Multimodal Foundation Model for  Low-Altitude Wireless Sensing}

\author{
Pengxuan~Gao,
Kai~Ying,~\IEEEmembership{Senior~Member,~IEEE},
Botao~Wu,
Jianhua~Mo,~\IEEEmembership{Senior~Member,~IEEE},
Qingsong~Wen,~\IEEEmembership{Senior~Member,~IEEE}

\thanks{Pengxuan~Gao, Kai~Ying, Botao~Wu, Jianhua~Mo are with the School of Electronic Information and Electrical Engineering, Shanghai Jiao Tong University, Shanghai 200240, China (e-mail: gpx123456789@sjtu.edu.cn; yingkai0301@sjtu.edu.cn; wubt2002@sjtu.edu.cn; mjh@sjtu.edu.cn).}
\thanks{Qingsong~Wen is with Squirrel Ai Learning, Bellevue, WA 98004 USA (e-mail: qingsongedu@gmail.com).}
\thanks{Kai~Ying is the corresponding author.}
}

\maketitle

\begin{abstract}
Low-altitude unmanned aerial vehicles (UAVs) are emerging as key platforms for wireless intelligence tasks. However, practical low-altitude wireless systems usually operate in complex urban environments, where visual occlusion, sparse geometric observations, multipath propagation, and sensor failures may degrade the reliability of single-modality models. To address these challenges, this paper proposes M3F-UAV, a missing-modality multimodal foundation model for low-altitude wireless sensing. The proposed framework learns a unified multimodal representation from visual, geometric, and wireless observations. Specifically, modality-specific pretrained feature extractors are adopted for RGB/depth images, LiDAR point clouds, and CSI matrices, respectively. Through cross-modal fusion and missing-modality-aware pretraining with feature-level masked reconstruction and UAV localization objectives, M3F-UAV can extract fixed-size features from different modality combinations and adapt them to downstream low-altitude wireless tasks with lightweight task heads. Experiments on the LAMBDA dataset show that M3F-UAV outperforms single-modality baselines and maintains robust performance under missing-modality settings.
\end{abstract}


\section{Introduction}
\IEEEPARstart{L}ow-altitude unmanned aerial vehicles (UAVs) have become an important component of emerging intelligent aerial systems, enabling applications such as urban inspection, emergency response, traffic monitoring, aerial communication, and low-altitude airspace management~\cite{1,gupta2016uav_networks,liu2025uav_localization_survey,low_altitude_networks_survey}. Compared with conventional ground terminals, UAVs operate in highly dynamic three-dimensional environments, where sensing and communication systems have to deal with complex urban layouts, visual occlusion, non-line-of-sight propagation, multipath reflection, and rapidly time-varying wireless channels~\cite{khuwaja2018uav_channel_survey,yan2023uav_urban_survey}. These characteristics make robust UAV perception and wireless intelligence essential for reliable low-altitude operations.

With the rapid development of artificial intelligence, many wireless sensing and communication tasks have been increasingly integrated with data-driven learning methods, including UAV localization~\cite{lidar_uav_localization,padhy2019uav_localization,wang2016deepfi}, beam prediction~\cite{charan2022drone_beam_prediction,salehi2020camera_beam,jiang2022lidar_beam,demirhan2022LiDAR_beam}, and CSI prediction~\cite{yang_channel_prediction,tong2018lstm_channel}. Existing studies have explored different single-modality solutions for these tasks. For instance, RGB images have been used for UAV localization and vision-assisted beam prediction~\cite{padhy2019uav_localization,salehi2020camera_beam}, LiDAR observations have been adopted for sensing-aided beam prediction~\cite{jiang2022lidar_beam,demirhan2022LiDAR_beam}, and CSI has been widely investigated for wireless localization and channel prediction~\cite{wang2016deepfi,tong2018lstm_channel,yang_channel_prediction}. Although these methods achieve promising results, their performance and robustness is inherently constrained by the sensing characteristics of the adopted modality.

To improve robustness, recent studies have explored multimodal sensing for wireless intelligence. DeepSense 6G provides synchronized multimodal sensing and communication data for joint sensing, communication, and positioning research~\cite{alkhateeb2022deepsense}. Camera images, position information, LiDAR, and GPS have also been fused for beam prediction in vehicular, UAV, and 6G communication scenarios~\cite{charan2022visionposition,ahmad2023visiondrone,tian2023multimodal,yeo2026multimodal,zheng2025m2beamllm}. In UAV communications, multimodal information such as UAV-captured images, transmitter/receiver positions, and communication settings has been used for channel prediction~\cite{xin2024uavchannel}. These works demonstrate the potential of multimodal fusion, but most of them are still designed for specific tasks.

With the success of foundation models represented by large language models such as GPT-3, PaLM, and LLaMA~\cite{brown2020gpt3,chowdhery2022palm,touvron2023llama}, the learning paradigm has gradually shifted from designing isolated task-specific models to pretraining general-purpose backbones for downstream adaptation. Inspired by this trend, wireless foundation models have recently attracted increasing attention. Some studies adapt pretrained language models to wireless tasks, such as TelecomGPT for telecom-domain language modeling~\cite{zou2024telecomgpt}, LLM4CP for CSI sequence prediction~\cite{liu2024llm4cp}, and LLM4WM for multi-task wireless modeling~\cite{liu2025llm4wm}. Other works develop wireless-native foundation models from wireless-domain data, such as LWM for universal channel embeddings~\cite{alikhani2024lwm}, WiFo for space-time-frequency CSI prediction~\cite{liu2024wifo}, AirFM-DDA for CSI estimation and prediction in the delay-Doppler-angle domain~\cite{bian2026airfmddaairinterfacefoundationmodel}, and WavesFM for unified wireless sensing, communication, and localization~\cite{aboulfotouh2025wavesfm}. Recent studies further explore CIR-CSI consistency for MIMO channel representation learning~\cite{jiang2025mimo}, environment-aware multi-task wireless foundation modeling~\cite{musefm}, and multimodal mixture-of-experts fusion for low-altitude ISAC networks~\cite{multimodalmoeisac}.

However, most existing wireless foundation models mainly focus on wireless-domain signals or predefined task settings, while missing-modality robustness in practical multimodal UAV sensing remains insufficiently explored. In real low-altitude systems, sensors may become unavailable due to occlusion, environmental degradation, hardware failure, or communication constraints. If a model cannot dynamically handle different modality combinations, a separate model has to be trained for each specific combination. As the number of modalities increases, the number of required models grows rapidly, which makes model maintenance, deployment, and task adaptation inefficient. Some recent multimodal foundation models, such as MultiMAE~\cite{bachmann2022multimae}, PhysioOmni~\cite{fu2025physioomni}, and X-Fi~\cite{chen2024xfi}, have shown the importance of flexible modality combinations and missing-modality modeling. Nevertheless, a unified multimodal foundation model for low-altitude wireless tasks under missing-modality conditions remains to be further investigated.

To address these limitations, we propose M3F-UAV, a Missing-Modality Multimodal Foundation Model for low-altitude wireless sensing. The proposed framework integrates RGB images, depth maps, LiDAR point clouds, and CSI matrices through modality-specific pretrained encoders, including a MobileNetV2-based \cite{MobileNetV2} encoder-decoder network for RGB/depth modalities, a Point Transformer encoder \cite{pointMAE} for LiDAR point clouds, and a Transformer-based encoder for antenna-domain CSI matrices. The extracted features are projected into a unified multimodal embedding space and fused by a cross-modal fusion backbone.

M3F-UAV is pretrained under missing-modality conditions. Specifically, stochastic modality dropping is used to simulate unavailable sensors, and one remaining modality is selected as the target modality for feature-level masked reconstruction. Together with the UAV localization objective, this strategy encourages the model to learn cross-modal correspondences, scene-level spatial semantics, and UAV-related sensing representations from incomplete multimodal observations. After pretraining, the backbone can extract fixed-size unified features from different modality combinations and can be transferred to UAV localization, beam prediction, and CSI prediction by attaching lightweight MLP-based task heads.

The main contributions of this paper are summarized as follows:

\begin{itemize}
    \item We propose M3F-UAV, a unified multimodal foundation model for low-altitude wireless tasks, which integrates RGB, depth, point-cloud, and CSI features through modality-specific pretrained encoders and a shared fusion backbone.

    \item We design a joint multimodal pretraining strategy that combines feature-level masked reconstruction and UAV localization, enabling the model to learn cross-modal feature correspondence and UAV-related spatial representations.

    \item We introduce a missing-modality-robust fusion mechanism based on stochastic modality dropping, target-modality masking, and adaptive token pooling, allowing the model to produce fixed-size fusion features from different available modality combinations.
\end{itemize}

\section{Proposed Method}

\begin{figure*}[t]
    \centering
    \includegraphics[width=\linewidth]{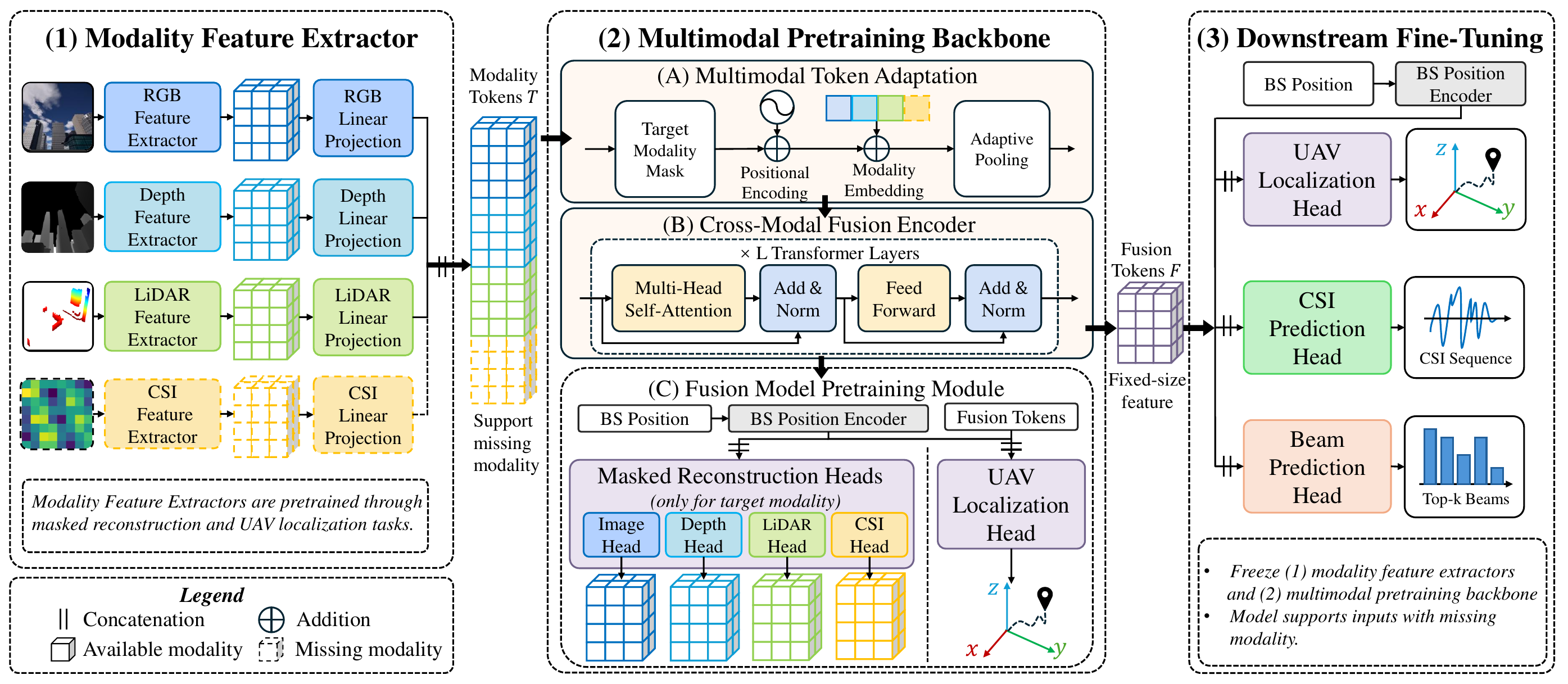}
\caption{
Overall architecture of the proposed multimodal foundation model for low-altitude wireless tasks. The framework contains three stages: modality feature extraction, multimodal pretraining, and downstream fine-tuning. The input modalities may be incomplete, and the dashed CSI branch is shown as an example of a missing modality.
}
\label{fig:framework}
\end{figure*}

\subsection{Overview of the Proposed Framework}

To learn transferable representations for UAV sensing under heterogeneous and incomplete sensing conditions, we propose a unified multimodal pretraining framework that maps visual, geometric, and wireless modalities into a shared representation space. As shown in Fig.~\ref{fig:framework}, the framework consists of three stages: modality-specific feature extraction, multimodal token adaptation and fusion pretraining, and downstream task fine-tuning. It supports both full-modality inputs and missing-modality scenarios.

Given a training sample, the multimodal input is denoted as
\begin{equation}
    \mathcal{X}=\{X^{r}, X^{d}, X^{p}, X^{c}, \mathbf{v}\},
\end{equation}
where \(X^{r}\), \(X^{d}\), \(X^{p}\), and \(X^{c}\) represent the RGB image, depth map, LiDAR point cloud, and CSI, respectively. The vector \(\mathbf{v}\) denotes the base-station position, including 3-D position and quaternion-based orientation. The complete modality set is
\begin{equation}
    \mathcal{M}=\{r,d,p,c\}.
\end{equation}

To improve robustness to incomplete observations, we introduce stochastic modality dropping during multimodal pretraining. A subset of modalities is randomly retained, denoted as \(\mathcal{S}\subseteq\mathcal{M}\). Only the retained modalities are processed by their corresponding feature extractors:
\begin{equation}
    \mathbf{Z}^{m}=\mathcal{E}_{m}(X^{m}), \quad m\in\mathcal{S},
\end{equation}
while discarded modalities are not fed into the multimodal backbone. This strategy simulates different missing-modality combinations and prevents the model from depending on complete modality inputs.

For each retained modality, the extractor outputs latent tokens
\begin{equation}
    \mathbf{Z}^{m}\in\mathbb{R}^{B\times N_m\times D_m},
\end{equation}
where \(B\), \(N_m\), and \(D_m\) denote the batch size, token number, and feature dimension, respectively. Since different modalities have heterogeneous token structures and feature dimensions, a modality-specific linear projection is introduced to map them into a unified dimension \(D\):
\begin{equation}
    \mathbf{T}^{m}
    =
    \mathbf{Z}^{m}\mathbf{W}_{m}+\mathbf{b}_{m},
    \quad
    \mathbf{T}^{m}\in\mathbb{R}^{B\times N_m\times D},
    \quad m\in\mathcal{S}.
\end{equation}
Here, \(\mathbf{W}_{m}\in\mathbb{R}^{D_m\times D}\) and \(\mathbf{b}_{m}\in\mathbb{R}^{D}\) are the learnable weight matrix and bias vector of the modality-specific projection for modality \(m\), respectively. The projected tokens are then processed by the multimodal token adaptation module, which applies target-modality masking, positional encoding, modality embedding, and adaptive pooling to obtain fixed-size aligned multimodal tokens.

After modality dropping, one retained modality \(t\in\mathcal{S}\) is selected as the reconstruction target. Its tokens are masked, while the remaining modalities serve as visible context:
\begin{equation}
    \mathcal{S}_{vis}=\mathcal{S}\setminus\{t\}.
\end{equation}
This enables the model to reconstruct a target modality from incomplete multimodal observations rather than from fixed full-modality inputs.

The adapted tokens are fed into a Transformer-based cross-modal fusion encoder, where self-attention captures complementary relationships among visual appearance, depth geometry, LiDAR point-cloud structure, and wireless propagation features. The resulting fusion tokens are optimized by two pretraining objectives: masked modality feature reconstruction and UAV localization. Instead of reconstructing raw modality data, such as images, point clouds, or CSI matrices, the reconstruction head predicts the latent feature tokens of the masked target modality. This feature-level reconstruction encourages cross-modal correspondence learning under missing-modality conditions, while the localization objective preserves spatially discriminative information in the fused representation.

After pretraining, the learned representation can be transferred to downstream tasks such as UAV localization, beam prediction, and CSI prediction by attaching lightweight task-specific heads. Since the model has observed diverse modality combinations during pretraining, it can adapt to different available modality sets during fine-tuning. The modality extractors and multimodal backbone can be frozen for efficient adaptation or partially fine-tuned when sufficient labeled data are available.

In addition, we consider modality failure during deployment. In practical UAV sensing systems, a modality may become unreliable due to hardware malfunction, occlusion, environmental interference, or adverse propagation conditions. To avoid corrupted inputs disturbing the fused representation, we introduce a Failure-Aware Modality Gating (FAMG) mechanism. A modality failure discriminator estimates the reliability of each modality based on pretrained modality features. Reliable modalities are retained for fusion, while failed modalities are removed. Benefiting from the missing-modality-aware design, the framework can perform robust inference even when some modalities are unavailable or unreliable.

\subsection{Modality Feature Extractors}

The goal of the modality feature extractors is to transform different raw sensing signals into compact token representations while preserving their modality-specific physical structures. Unlike a single shared encoder, we use separate backbones for different modalities because RGB/depth images, LiDAR point clouds, and CSI measurements exhibit fundamentally different data organizations.

\subsubsection{RGB and Depth Feature Extractors}

\begin{figure*}[t]
    \centering
    \includegraphics[width=\linewidth]{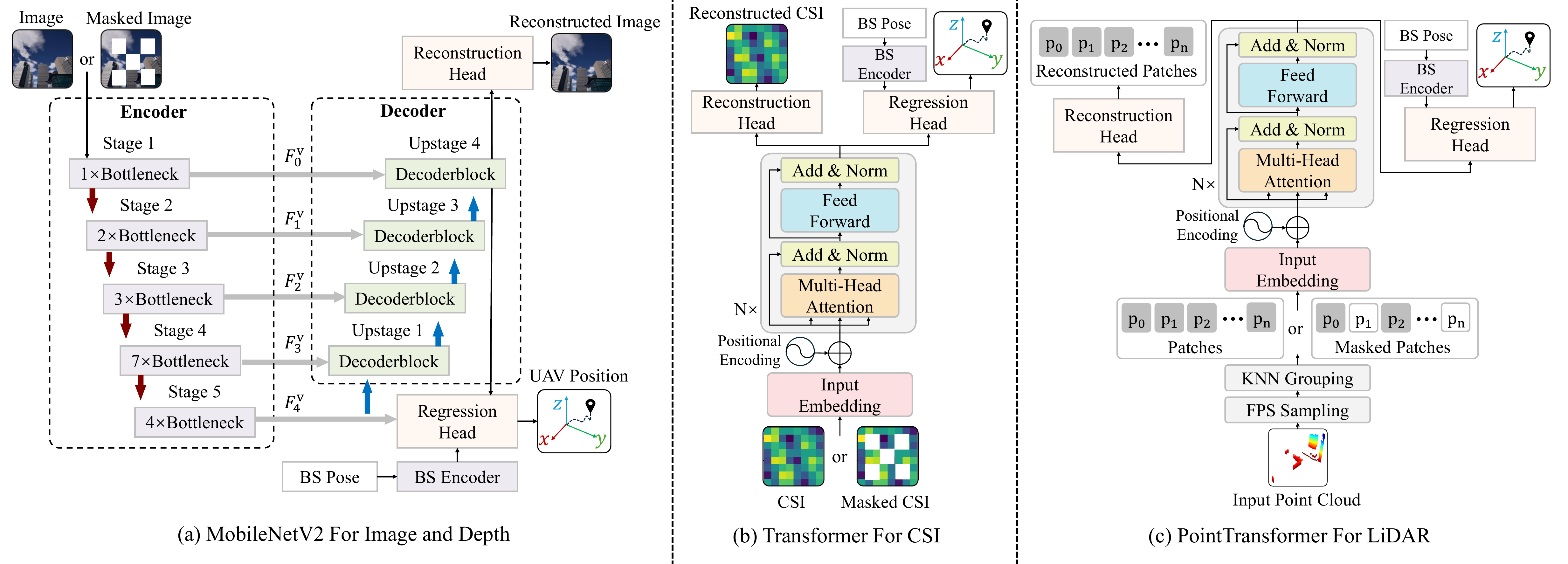}
\caption{Network architectures used for pretraining the image, depth, CSI, and LiDAR modalities. (a) MobileNetV2 for image and depth data, (b) Transformer for CSI data, and (c) Point Transformer for LiDAR point clouds. All modality-specific models are pretrained with reconstruction and UAV localization tasks.}
\label{fig:framework_pre}
\end{figure*}

For the visual modalities, we adopt a lightweight convolutional encoder-decoder architecture for both RGB and depth feature extraction. Convolutional neural networks are adopted because they efficiently capture local spatial patterns and hierarchical visual representations with relatively low computational complexity. As shown in Fig.~\ref{fig:framework_pre} (a), The visual branch consists of a MobileNetV2-style encoder with inverted residual blocks and a U-Net-like decoder with skip connections, enabling efficient multi-scale spatial feature learning from dense visual observations.

Given a visual input
\begin{equation}
    X^{v}\in\mathbb{R}^{B\times C_v\times H\times W},
    \quad v\in\{r,d\},
\end{equation}
where \(r\) and \(d\) denote RGB and depth modalities, respectively. Here, \(C_r=3\) for RGB images and \(C_d=1\) for single-channel depth maps.The input is first processed by a convolutional stem:
\begin{equation}
    F_{0}^{v}
    =
    \mathrm{ReLU6}
    \left(
    \mathrm{BN}
    \left(
    \mathrm{Conv}_{3\times3}(X^{v})
    \right)
    \right).
\end{equation}
The resulting feature is then fed into a sequence of MobileNetV2 inverted residual (IR) stages. Each inverted residual block follows an expansion-depthwise-projection structure:
\begin{equation}
    \mathrm{IR}(F)
    =
    \mathrm{Conv}_{1\times1}^{proj}
    \left(
    \mathrm{DWConv}_{3\times3}
    \left(
    \mathrm{Conv}_{1\times1}^{exp}(F)
    \right)
    \right),
\end{equation}
where \(\mathrm{DWConv}_{3\times3}\) denotes the \(3\times3\) depthwise convolution (DWConv). A residual connection is used when the stride is one and the input and output dimensions are identical:
\begin{equation}
    F_{out}=F+\mathrm{IR}(F).
\end{equation}
This design reduces computational cost while maintaining sufficient representation capacity for visual UAV sensing.

The encoder produces hierarchical features:
\begin{equation}
    \{F_{0}^{v},F_{1}^{v},F_{2}^{v},F_{3}^{v},F_{4}^{v}\}
    =
    \mathcal{E}_{v}(X^{v}),
\end{equation}
where shallow features preserve local spatial details and the deepest feature \(F_{4}^{v}\) captures high-level semantic information.

The visual encoder is pretrained with two objectives: block-level masked reconstruction and UAV coordinate regression. For masked reconstruction, the input image or depth map is divided into non-overlapping blocks, and a subset of blocks is randomly masked in the pixel space. A U-Net-style decoder reconstructs the original visual input by progressively upsampling the deepest feature and fusing it with the corresponding encoder features:
\begin{equation}
    D_{i}^{v}
    =
    \phi_{i}
    \left(
    \left[
    \mathrm{Up}(D_{i+1}^{v});
    \overline{F}_{i}^{v}
    \right]
    \right),
\end{equation}
where \(\mathrm{Up}(\cdot)\) denotes bilinear upsampling, \([\cdot;\cdot]\) denotes channel-wise concatenation, and \(\phi_i(\cdot)\) is a convolutional refinement block. The reconstruction head predicts \(\widehat{X}^{v}\), encouraging the encoder to infer missing visual content from spatial context and improving robustness to occlusion and invalid depth regions.

For UAV coordinate regression, the deepest feature \(F_{4}^{v}\) is projected into a compact feature map \(G_{deep}^{v}\), while the decoder provides a high-resolution feature \(G_{dec}^{v}\). These two features are aligned in resolution, concatenated, and fused:
\begin{equation}
    G_{fuse}^{v}
    =
    \phi_{fuse}
    \left(
    \left[
    G_{dec}^{v};
    \mathrm{Up}(G_{deep}^{v})
    \right]
    \right),
\end{equation}
where \(\phi_{fuse}(\cdot)\) denotes a convolutional fusion block. The fused feature is then converted into a compact visual representation by global average pooling (GAP):
\begin{equation}
    \mathbf{f}^{v}
    =
    \mathrm{GAP}(G_{fuse}^{v}).
\end{equation}
Here, \(\mathrm{GAP}(\cdot)\) averages each feature channel over the spatial dimensions, thereby producing a fixed-dimensional feature vector for the visual modality.
Meanwhile, the base-station position vector \(\mathbf{v}_{\mathrm{BS}}\) is encoded by a lightweight MLP:
\begin{equation}
    \mathbf{q}
    =
    \mathcal{E}_{BS}(\mathbf{v}_{BS}).
\end{equation}
The visual feature and base-station position feature are concatenated and fed into an MLP regression head:
\begin{equation}
    \widehat{\mathbf{y}}^{v}
    =
    \mathcal{F}_{reg}^{v}
    \left(
    [\mathbf{f}^{v};\mathbf{q}]
    \right),
    \quad
    \widehat{\mathbf{y}}^{v}\in\mathbb{R}^{3}.
\end{equation}

After pretraining, the reconstruction decoder and regression head are removed, and the MobileNetV2 encoder is retained as the visual feature extractor. For multimodal fusion, the deepest feature map \(F_{4}^{v}\) is projected into the unified multimodal embedding space:
\begin{equation}
    \mathbf{T}^{v}
    =
    \mathbf{F}_{4}^{v}\mathbf{W}_{p}
    +
    \mathbf{b}_{v},
\end{equation}
where \(\mathbf{T}^{v}\) is used as the visual token input to the multimodal token adaptation module.

\subsubsection{CSI Feature Extractor}

As shown in Fig.~\ref{fig:framework_pre}(b), for the wireless modality, we construct a Transformer-based CSI feature extractor to process the complex-valued channel matrix between the transmit and receive antenna arrays. Different from wideband CSI with an explicit subcarrier dimension, the CSI in this work is calculated only at the center frequency and is therefore represented as a single complex channel matrix. The Transformer architecture is adopted because its self-attention mechanism can effectively capture global dependencies and spatial correlations among antenna-domain CSI elements.

Let the CSI matrix between the UAV transmitter and the base-station receiver be
\begin{equation}
    H^{c}\in\mathbb{C}^{N_r\times N_t},
\end{equation}
where \(N_t\) denotes the number of transmit antennas at the UAV, and \(N_r\) denotes the number of receive antennas at the base station. Each element \(H^{c}_{i,j}\) represents the complex channel coefficient from the \(j\)-th UAV transmit antenna to the \(i\)-th base-station receive antenna. To enable real-valued neural network processing, we concatenate the real and imaginary parts along the feature dimension:
\begin{equation}
    X^{c}
    =
    \left[
    \mathrm{Re}(H^{c});
    \mathrm{Im}(H^{c})
    \right]
    \in\mathbb{R}^{N_r\times 2N_t}.
\end{equation}
For a batch of samples,
\begin{equation}
    X^{c}\in\mathbb{R}^{B\times N_r\times 2N_t}.
\end{equation}
In this representation, each row corresponds to one receive antenna, while the feature dimension contains the real and imaginary channel coefficients associated with the transmit antenna array. This preserves amplitude-phase information without requiring complex-valued network operations.

The CSI matrix is then converted into antenna-domain tokens by projecting each receive-antenna feature vector into a latent space:
\begin{equation}
    \mathbf{H}^{c}_{0}
    =
    X^{c}\mathbf{W}_{c}+\mathbf{b}_{c},
    \quad
    \mathbf{H}^{c}_{0}\in\mathbb{R}^{B\times N_r\times D_c},
\end{equation}
where \(D_c\) is the CSI token dimension. A learnable antenna positional embedding is added to preserve antenna-domain ordering:
\begin{equation}
    \mathbf{H}^{c}_{0}
    =
    \mathbf{H}^{c}_{0}
    +
    \mathbf{E}^{c}_{ant}.
\end{equation}
The resulting CSI tokens are processed by a Transformer encoder:
\begin{equation}
    \mathbf{Z}^{c}
    =
    \mathrm{TransformerEncoder}_{c}
    \left(
    \mathbf{H}^{c}_{0}
    \right).
\end{equation}
The output tokens \(\mathbf{Z}^{c}\) are used for CSI reconstruction and downstream wireless tasks.

The CSI encoder is pretrained with two objectives: masked CSI reconstruction and UAV coordinate regression. For masked reconstruction, we perform feature-channel masking on the concatenated real-imaginary CSI representation. A subset of feature channels is randomly masked across the antenna dimension, forcing the model to infer missing channel coefficients from the remaining observations. Let
\begin{equation}
    \mathbf{M}^{c}
    \in
    \{0,1\}^{B\times N_r\times 2N_t}
\end{equation}
denote the CSI mask, where 1 indicates masked entries. The masked input is formulated as
\begin{equation}
    \overline{X}^{c}
    =
    X^{c}\odot(1-\mathbf{M}^{c})
    +
    \nu_{mask}\mathbf{M}^{c},
\end{equation}
where \(\nu_{mask}\) is the mask value. The masked input is encoded by the CSI Transformer, and a lightweight decoder reconstructs the original real-valued CSI matrix:
\begin{equation}
    \widehat{X}^{c}
    =
    \mathcal{D}_{c}
    \left(
    \mathbf{Z}^{c}
    \right).
\end{equation}
This objective encourages the encoder to capture correlations among antenna responses and learn robust wireless representations from incomplete channel observations.

For UAV coordinate regression, the CSI tokens are first aggregated into a compact feature representation:
\begin{equation}
    \mathbf{f}^{c}
    =
    \mathrm{GAP}
    \left(
    \mathbf{Z}^{c}
    \right).
\end{equation}
The base-station position vector \(\mathbf{v}_{\mathrm{BS}}\) is encoded by a lightweight MLP:
\begin{equation}
    \mathbf{q}
    =
    \mathcal{E}_{BS}(\mathbf{v}_{BS}).
\end{equation}
The CSI feature and base-station position feature are concatenated and fed into an MLP regression head:
\begin{equation}
    \widehat{\mathbf{y}}^{c}
    =
    \mathcal{F}_{reg}^{c}
    \left(
    [
    \mathbf{f}^{c};
    \mathbf{q}
    ]
    \right),
    \quad
    \widehat{\mathbf{y}}^{c}\in\mathbb{R}^{3}.
\end{equation}

After pretraining, the reconstruction decoder and regression head are removed, and the CSI Transformer encoder is retained as the wireless feature extractor. For multimodal fusion, \(\mathbf{Z}^{c}\) is projected into the unified multimodal embedding space:
\begin{equation}
    \mathbf{T}^{c}
    =
    \mathbf{Z}^{c}\mathbf{W}_{p}^{c}
    +
    \mathbf{b}_{p}^{c},
\end{equation}
where \(\mathbf{T}^{c}\) denotes the projected CSI feature used as the input of the multimodal token adaptation module.
\subsubsection{LiDAR Point Cloud Feature Extractor}

As shown in Fig.~\ref{fig:framework_pre}(c), for the LiDAR modality, the input is an unordered 3-D point cloud:
\begin{equation}
    X^{p}=\{\mathbf{x}_{i}\}_{i=1}^{N},
    \quad
    \mathbf{x}_{i}\in\mathbb{R}^{3},
\end{equation}
where \(N\) denotes the number of LiDAR points. Since raw point clouds are irregular and unordered, we first partition them into local geometric patches using farthest point sampling (FPS) and \(K\)-nearest-neighbor (KNN) grouping. FPS selects \(G\) representative centers:
\begin{equation}
    \{\mathbf{s}_{g}\}_{g=1}^{G}
    =
    \mathrm{FPS}(X^{p}),
\end{equation}
and KNN constructs a local patch around each center:
\begin{equation}
    \mathcal{P}_{g}
    =
    \mathrm{KNN}(\mathbf{s}_{g}, X^{p}, K),
\end{equation}
where \(K\) is the number of neighboring points. Each patch is represented by relative coordinates:
\begin{equation}
    \widetilde{\mathcal{P}}_{g}
    =
    \{\mathbf{x}_{i}-\mathbf{s}_{g}\mid \mathbf{x}_{i}\in\mathcal{P}_{g}\}.
\end{equation}

Each local patch is embedded into a patch token through a point-wise MLP followed by max pooling:
\begin{equation}
    \mathbf{t}^{p}_{g}
    =
    \mathrm{MaxPool}
    \left(
    \mathrm{MLP}_{p}(\widetilde{\mathcal{P}}_{g})
    \right).
\end{equation}
The sampled center is further encoded as a positional embedding and added to the patch token:
\begin{equation}
    \mathbf{h}^{p}_{g}
    =
    \mathbf{t}^{p}_{g}
    +
    \phi_{p}(\mathbf{s}_{g}),
\end{equation}
where \(\phi_{p}(\cdot)\) is a lightweight MLP. The resulting point tokens are stacked as
\begin{equation}
    \mathbf{H}^{p}_{0}
    =
    [\mathbf{h}^{p}_{1},\mathbf{h}^{p}_{2},\ldots,\mathbf{h}^{p}_{G}]
    \in\mathbb{R}^{B\times G\times D_p}.
\end{equation}

To capture long-range geometric dependencies, the point tokens are processed by a Point Transformer encoder with relative positional encoding between patch centers:
\begin{equation}
    \mathbf{Z}^{p}
    =
    \mathrm{TransformerEncoder}_{p}
    \left(
    \mathbf{H}^{p}_{0},
    \{\mathbf{s}_{g}\}_{g=1}^{G}
    \right).
\end{equation}
The encoded tokens \(\mathbf{Z}^{p}\) preserve both local shape patterns and global geometric layouts of the LiDAR observation.

The LiDAR encoder is pretrained with masked local patch reconstruction and UAV coordinate regression. For reconstruction, a subset of point tokens is randomly masked and replaced by learnable mask tokens in the decoder. The reconstruction head predicts the local coordinates of masked patches:
\begin{equation}
    \widehat{\mathcal{P}}_{g}
    =
    \mathcal{D}_{p}(\mathbf{z}^{p}_{g}),
    \quad
    g\in\Omega_{mask},
\end{equation}
where \(\Omega_{mask}\) denotes the masked patch indices. The reconstruction loss is computed using Chamfer distance.

For coordinate regression, mean-pooled and max-pooled point features are concatenated with the base-station position embedding:
\begin{equation}
    \mathbf{f}^{p}
    =
    \left[
    \mathrm{MeanPool}(\mathbf{Z}^{p});
    \mathrm{MaxPool}(\mathbf{Z}^{p});
    \mathcal{E}_{BS}(\mathbf{v}_{BS})
    \right],
\end{equation}
and fed into an MLP head:
\begin{equation}
    \widehat{\mathbf{y}}^{p}
    =
    \mathcal{F}_{reg}^{p}(\mathbf{f}^{p}),
    \quad
    \widehat{\mathbf{y}}^{p}\in\mathbb{R}^{3}.
\end{equation}

After pretraining, the reconstruction decoder and regression head are removed, and the Point Transformer encoder is retained as the LiDAR feature extractor. For multimodal fusion, \(\mathbf{Z}^{p}\) is projected into the unified multimodal embedding space:
\begin{equation}
    \mathbf{T}^{p}
    =
    \mathbf{Z}^{p}\mathbf{W}_{p}
    +
    \mathbf{b}_{p},
\end{equation}
where \(\mathbf{T}^{p}\) denotes the projected point-cloud feature used as the input of the multimodal token adaptation module.

\subsection{Multimodal Pretraining Backbone}

After obtaining modality-specific representations, the multimodal pretraining backbone is introduced to learn a unified representation across visual, geometric, and wireless observations. As illustrated in Fig.~\ref{fig:framework}, the backbone consists of three components: multimodal token adaptation, cross-modal fusion encoder, and fusion model pretraining module. The key objective of this backbone is to learn transferable multimodal features under incomplete modality conditions. Therefore, the input of this stage is not necessarily a complete set of modality features. Instead, it is a retained modality set \(\mathcal{S}\subseteq\mathcal{M}\), which is generated by stochastic modality dropping during pretraining.

\subsubsection{Multimodal Token Adaptation}

The modality feature extractors produce heterogeneous representations with different feature dimensions and token organizations. In addition, due to stochastic modality dropping, the input of the multimodal token adaptation module may contain only a subset of the complete modality set. Therefore, the token adaptation module is designed to process incomplete modality features and transform the resulting variable-length multimodal sequence into a fixed-length token representation.

Let the complete modality set be denoted as
\begin{equation}
    \mathcal{M}=\{r,d,p,c\},
\end{equation}
where \(r\), \(d\), \(p\), and \(c\) represent RGB, depth, LiDAR point cloud, and CSI modalities, respectively. During multimodal pretraining, stochastic modality dropping is first applied to generate a retained modality set
\begin{equation}
    \mathcal{S}\subseteq\mathcal{M}.
\end{equation}
Only the modalities in \(\mathcal{S}\) are processed by their corresponding feature extractors. For each retained modality \(m\in\mathcal{S}\), the unified multimodal feature map is denoted as \(\mathbf{T}^{m}\). Positional embeddings and modality embeddings are then added to the projected tokens:
\begin{equation}
    \mathbf{A}^{m}
    =
    \mathbf{T}^{m}
    +
    \mathbf{E}_{pos}^{m}
    +
    \mathbf{E}_{mod}^{m},
    \quad
    m\in\mathcal{S}.
\end{equation}
Here, \(\mathbf{E}_{pos}^{m}\) encodes the internal token position of modality \(m\), and \(\mathbf{E}_{mod}^{m}\) is a learnable modality embedding that identifies the sensing source.

After embedding, target-modality masking is performed for feature-level reconstruction. It should be emphasized that stochastic modality dropping and target-modality masking are two different operations. The former determines the retained modality set \(\mathcal{S}\), while the latter selects one modality from \(\mathcal{S}\) as the reconstruction target. Each retained modality has the same probability of being selected:
\begin{equation}
    P(t=m)=\frac{1}{|\mathcal{S}|},
    \quad
    m\in\mathcal{S}.
\end{equation}
The selected modality \(t\) is masked by replacing its content tokens with learnable mask tokens, while its positional and modality embeddings are still retained:
\begin{equation}
    \overline{\mathbf{A}}^{t}
    =
    \mathbf{M}_{t}
    +
    \mathbf{E}_{pos}^{t}
    +
    \mathbf{E}_{mod}^{t},
\end{equation}
where \(\mathbf{M}_{t}\in\mathbb{R}^{B\times N_t\times D}\) denotes the learnable mask token sequence of the target modality. The remaining retained modalities are used as visible contextual modalities:
\begin{equation}
    \mathcal{S}_{vis}
    =
    \mathcal{S}\setminus\{t\}.
\end{equation}
The multimodal token sequence before adaptive pooling is then formulated as
\begin{equation}
    \mathbf{A}_{var}
    =
    \left[
    \{\mathbf{A}^{m}\}_{m\in\mathcal{S}_{vis}};
    \overline{\mathbf{A}}^{t}
    \right].
\end{equation}

Due to missing modalities, the length of \(\mathbf{A}_{var}\) is not fixed. Thus, different modality combinations lead to different input lengths for the multimodal fusion encoder. To obtain a consistent input size for subsequent cross-modal fusion, we apply an adaptive token pooling operation after modality embedding and target-modality masking:
\begin{equation}
    \mathbf{A}
    =
    \mathrm{ATP}
    \left(
    \mathbf{A}_{var}
    \right),
    \quad
    \mathbf{A}
    \in
    \mathbb{R}^{B\times K_a\times D},
\end{equation}
where \(K_a\) is the predefined number of adapted multimodal tokens. Different from using pooling to balance the token length of each individual modality, the adaptive token pooling operation is applied to the concatenated multimodal sequence. Its purpose is to convert the variable-length token sequence caused by missing-modality inputs into a fixed-length representation.

The resulting adapted tokens \(\mathbf{A}\) contain information from the available visible modalities and the masked target modality. Since \(\mathbf{A}\) has a fixed shape regardless of the retained modality set \(\mathcal{S}\), the subsequent cross-modal fusion encoder can use the same architecture for full-modality and missing-modality inputs.

\subsubsection{Cross-Modal Fusion Encoder}

After multimodal token adaptation, the variable-length multimodal sequence caused by missing modalities has been converted into a fixed-length representation \(\mathbf{A}\). Since \(\mathbf{A}\) has a consistent size for different modality combinations, it can be directly fed into the cross-modal fusion encoder without additional token concatenation or modality-specific padding.

To model the interactions among heterogeneous sensing modalities, we employ a multi-layer Transformer encoder as the cross-modal fusion encoder. The input of the first layer is defined as
\begin{equation}
    \mathbf{F}_{0}
    =
    \mathbf{A}.
\end{equation}
Then, the adapted multimodal tokens are processed by \(L\) stacked cross-modal fusion layers:
\begin{equation}
    \mathbf{F}_{\ell}
    =
    \mathrm{CFE}_{\ell}
    \left(
    \mathbf{F}_{\ell-1}
    \right),
    \quad
    \ell=1,2,\ldots,L,
\end{equation}
where \(\mathrm{CFE}_{\ell}(\cdot)\) denotes the \(\ell\)-th cross-modal fusion layer. Each layer consists of multi-head self-attention, a feed-forward network, residual connections, and layer normalization. The multi-head self-attention operation allows each adapted token to attend to all other tokens, thereby establishing global dependencies among the available visual, geometric, and wireless features.

The output of the final fusion layer is used as the unified multimodal representation:
\begin{equation}
    \mathbf{F}
    =
    \mathbf{F}_{L}
    \in
    \mathbb{R}^{B\times K_a\times D}.
\end{equation}
Since both \(K_a\) and \(D\) are fixed, the fused representation \(\mathbf{F}\) has a consistent size regardless of which modalities are available. This property enables the same fusion backbone and downstream task heads to be applied to full-modality and missing-modality scenarios. A compact global feature can be further obtained by average pooling over the fused tokens:
\begin{equation}
    \mathbf{f}
    =
    \mathrm{AvgPool}
    \left(
    \mathbf{F}
    \right)
    \in
    \mathbb{R}^{B\times D}.
\end{equation}
The token-level feature \(\mathbf{F}\) is used for feature-level masked reconstruction, while the pooled feature \(\mathbf{f}\) is used for UAV coordinate regression.

\subsubsection{Fusion Model Pretraining Module}

The fusion model pretraining module is designed to make the multimodal backbone learn both cross-modal correspondence and task-relevant spatial information. It contains two pretraining tasks: feature-level masked reconstruction and UAV coordinate regression. Different from the single-modality reconstruction tasks that recover raw modality inputs, the multimodal masked reconstruction task reconstructs the latent feature of the selected target modality. This design avoids forcing the fusion backbone to generate low-level modality-specific signals and instead encourages it to learn semantic, geometric, and wireless correspondences across modalities.

For the feature-level masked reconstruction task, the target modality \(t\) is first selected from the retained modality set \(\mathcal{S}\). Its unmasked adapted feature \(\widetilde{\mathbf{T}}^{t}\) is used as the reconstruction target, while its input content in the fusion encoder is replaced by mask tokens. Given the fusion tokens \(\mathbf{F}\), a target-specific reconstruction head predicts the feature tokens of the target modality:
\begin{equation}
    \widehat{\mathbf{T}}^{t}
    =
    \mathcal{R}_{t}(\mathbf{F}),
    \quad
    \widehat{\mathbf{T}}^{t}
    \in
    \mathbb{R}^{B\times K_t\times D}.
\end{equation}
The reconstruction loss is defined in the latent feature space:
\begin{equation}
    \mathcal{L}_{feat}
    =
    \frac{1}{B K_t D}
    \left\|
    \widehat{\mathbf{T}}^{t}
    -
    \widetilde{\mathbf{T}}^{t}
    \right\|_{2}^{2}.
\end{equation}
This objective requires the fusion encoder to infer the target modality representation from the remaining visible modalities.

The second pretraining task is UAV coordinate regression. The pooled fusion representation \(\mathbf{f}\) is concatenated with the base-station position feature to predict the 3-D UAV position:
\begin{equation}
    \mathbf{q}
    =
    \mathcal{E}_{BS}(\mathbf{v}_{BS}),
\end{equation}
\begin{equation}
    \widehat{\mathbf{y}}
    =
    \mathcal{F}_{loc}
    \left(
    [
    \mathbf{f};
    \mathbf{q}
    ]
    \right),
    \quad
    \widehat{\mathbf{y}}\in\mathbb{R}^{3},
\end{equation}
where \(\mathbf{v}_{\mathrm{BS}}\) denotes the base-station position vector, \(\mathbf{q}\) is the base-station position embedding, and \(\mathcal{F}_{loc}(\cdot)\) is an MLP localization head. The localization loss is written as
\begin{equation}
    \mathcal{L}_{loc}
    =
    \left\|
    \widehat{\mathbf{y}}
    -
    \mathbf{y}
    \right\|_{2}^{2},
\end{equation}
where \(\mathbf{y}\in\mathbb{R}^{3}\) is the ground-truth UAV position.

Through this pretraining strategy, the fusion backbone learns to aggregate incomplete multimodal inputs, reconstruct missing target-modality features, and preserve UAV-related spatial information. As a result, after pretraining, the model can be directly adapted to downstream tasks with different modality combinations.

\subsection{Downstream Fine-Tuning}

After multimodal pretraining, the modality feature extractors, multimodal token adaptation module, and cross-modal fusion encoder are retained as the pretrained backbone. With stochastic modality dropping and adaptive token pooling, the backbone can produce fixed-size representations from different modality combinations. Given an available modality set \(\mathcal{S}_{ft}\subseteq\mathcal{M}\), the fused representation is obtained as
\begin{equation}
    \mathbf{F}
    =
    \mathcal{B}
    \left(
    \{X^{m}\}_{m\in\mathcal{S}_{ft}}
    \right)
    \in
    \mathbb{R}^{B\times K_a\times D},
\end{equation}
where \(\mathcal{B}(\cdot)\) denotes the pretrained multimodal backbone. Since the shape of \(\mathbf{F}\) is independent of the available modality set, the same downstream head can be used for both full-modality and missing-modality inputs.

For downstream adaptation, all task heads are implemented as lightweight MLPs, allowing the pretrained representation to be evaluated without introducing complex task-specific designs. We adopt a two-stage fine-tuning strategy: the backbone is first frozen and only the task head is trained, and then the backbone is unfrozen and jointly optimized with a smaller learning rate. This stabilizes early fine-tuning while enabling task-specific adaptation.

\subsection{Failure-Aware Modality Gating}

\begin{figure}[t]
    \centering
    \includegraphics[width=\linewidth]{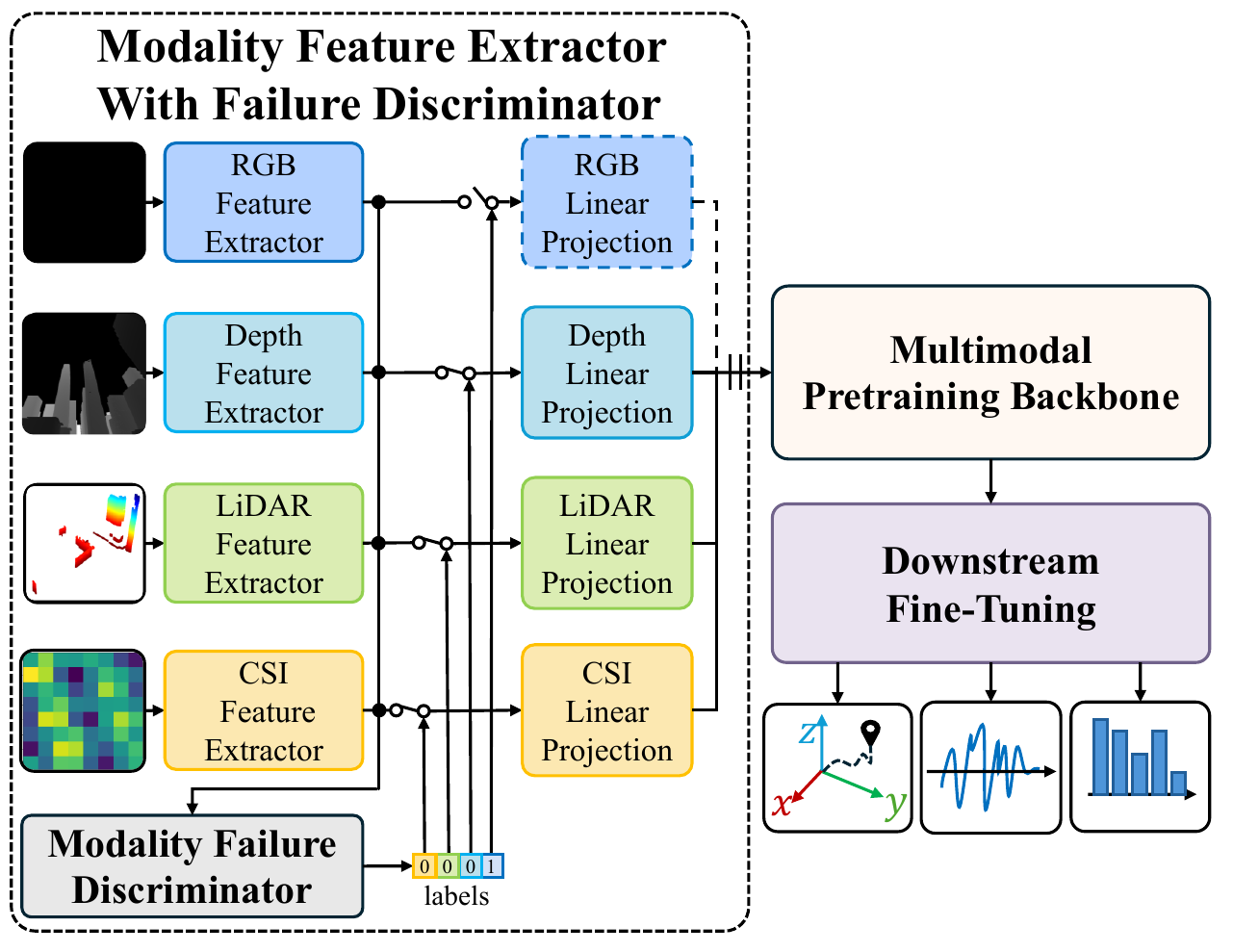}
\caption{
Overall architecture of the proposed Failure-Aware Modality Gating model. The dashed RGB branch is shown as an example of a failed modality.
}
\label{fig:failure_discriminator}
\end{figure}

In addition to missing-modality training, we further consider modality failure during practical deployment. In real UAV-related wireless systems, a sensor modality may become unreliable due to hardware malfunction, visual occlusion, environmental interference, or adverse propagation conditions. Directly feeding failed modality features into the multimodal fusion backbone may disturb the fused representation and degrade downstream performance. Therefore, we introduce a Failure-Aware Modality Gating (FAMG) module before multimodal fusion, as illustrated in Fig.~\ref{fig:failure_discriminator}.

The proposed FAMG module is built upon the pretrained modality feature extractors. Specifically, after the modality-specific encoders are trained, an additional Modality Failure Discriminator (MFD) is trained to classify whether the input of each modality is normal or failed. In this work, a simple failure simulation strategy is adopted. For RGB and depth modalities, modality failure is simulated by fully occluding the input image or depth map to simulate the night or dark scenario. For LiDAR point clouds and CSI matrices, severe noise is added to simulate unreliable geometric and wireless observations. The MFD takes the modality features extracted by the pretrained encoders as input and predicts a binary reliability label:
\begin{equation}
    s_m = \mathcal{D}_m(\mathbf{Z}^{m}), \quad m\in\mathcal{M},
\end{equation}
where \(\mathcal{D}_m(\cdot)\) denotes the failure discriminator for modality \(m\), and \(s_m\in\{0,1\}\) indicates whether the modality is reliable or failed.

During inference, if a modality is predicted as failed, its feature is removed from the input feature set before being sent to the multimodal token adaptation module:
\begin{equation}
    \mathcal{S}_{rel}
    =
    \{m\mid s_m=0, m\in\mathcal{M}\},
\end{equation}
where \(\mathcal{S}_{rel}\) denotes the set of reliable modalities. Only the features of reliable modalities are projected into the unified multimodal embedding space and processed by the pretrained multimodal backbone. Since M3F-UAV is pretrained with stochastic modality dropping and supports missing-modality inputs, no additional modification is required when failed modalities are actively suppressed. This design prevents corrupted modality features from degrading the fusion representation and improves the robustness of the model under practical sensor failure conditions.

\section{Experimental Setup and Result}

\subsection{Dataset Description}
\subsubsection{Dataset Overview}
The dataset used in this work is LAMBDA (Low-Altitude Multi-modal Base DAtaset), a comprehensive dataset that fuses fine-grained electromagnetic simulations with cinematic-quality visual rendering. It is specifically tailored for low-altitude UAV sensing scenarios. Different from conventional single-modality datasets, LAMBDA provides synchronized visual, geometric, inertial, and electromagnetic observations.The dataset used in this work is publicly available\footnote{\url{https://www.lambda6g.net/}}.



\subsubsection{Sensor Modalities and Configurations}
LAMBDA contains complementary visual, geometric, and wireless sensing modalities, as summarized in Table~\ref{tab:lambda_sensors}. The visual sensors provide RGB images and depth maps with a resolution of \(1920\times1080\), a \(100^{\circ}\) field of view, and a frame rate of 60 Hz. The LiDAR sensor generates 3-D point clouds with 512 scanning lines, a sensing range of 150 m, a horizontal field of view from \(-60^{\circ}\) to \(+60^{\circ}\), a vertical field of view from \(0^{\circ}\) to \(+90^{\circ}\), and a scan rate of 20 Hz.

For the wireless modality, LAMBDA provides measurements over multiple frequency bands, including 3.5 GHz, 4.9 GHz, 7 GHz, and 28 GHz. In this paper, the 4.9 GHz CSI is used for wireless-domain modeling, together with propagation parameters such as path delay \(\tau\), Doppler shift, AoD, and AoA. Since different sensors have different sampling rates and measurement mechanisms, LAMBDA applies spatio-temporal synchronization to align RGB images, depth maps, LiDAR point clouds, IMU data, CSI data, and FMCW LiDAR measurements using timestamps and spatial poses.

\begin{table}[t]
    \centering
    \caption{Main sensor configurations of LAMBDA.}
    \label{tab:lambda_sensors}
    \begin{tabular}{c|c}
        \hline
        \textbf{Sensor} & \textbf{Configuration} \\
        \hline
        Camera & \(1920\times1080\), \(100^{\circ}\) FOV, 60 Hz \\
        LiDAR & 512 lines, 150 m, 20 Hz \\
        Wireless & 4.9 GHz, CSI, \(\tau\), Doppler, AoD/AoA \\
        \hline
    \end{tabular}
\end{table}

\subsubsection{Urban Scenario Subset}

In this work, experiments are conducted on the urban scenario subset of LAMBDA. This subset covers diverse urban layouts, including high-rise urban canyons, medium-density residential blocks, and open squares or harbor areas, providing complex propagation and perception conditions with blockage, multipath, visual occlusion, and geometric sparsity.

\begin{figure}[t]
    \centering
    \subfloat[Scene 1]{
        \includegraphics[width=0.48\linewidth]{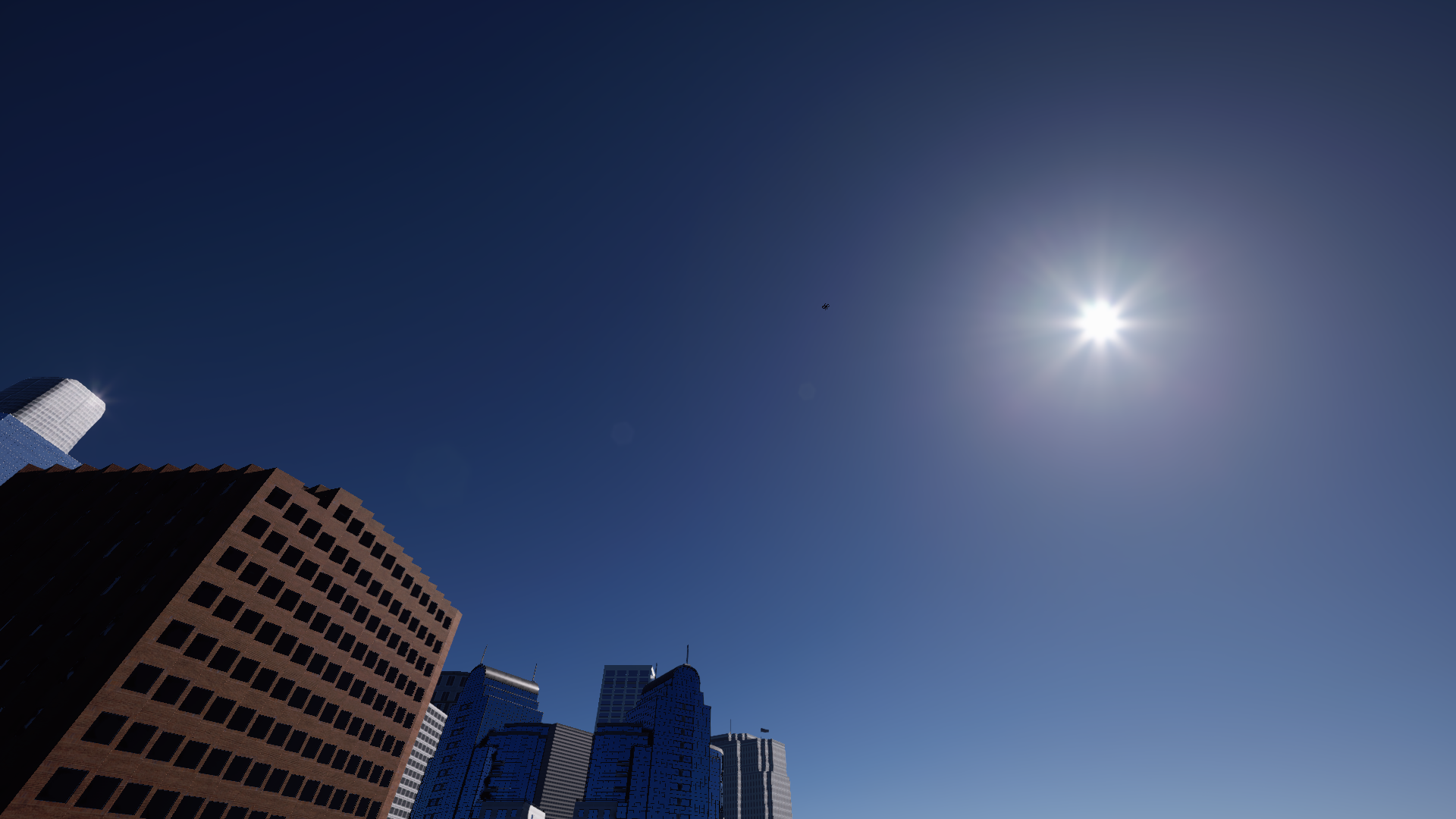}
        \label{fig:scenario1}
    }
    \subfloat[Scene 2]{
        \includegraphics[width=0.48\linewidth]{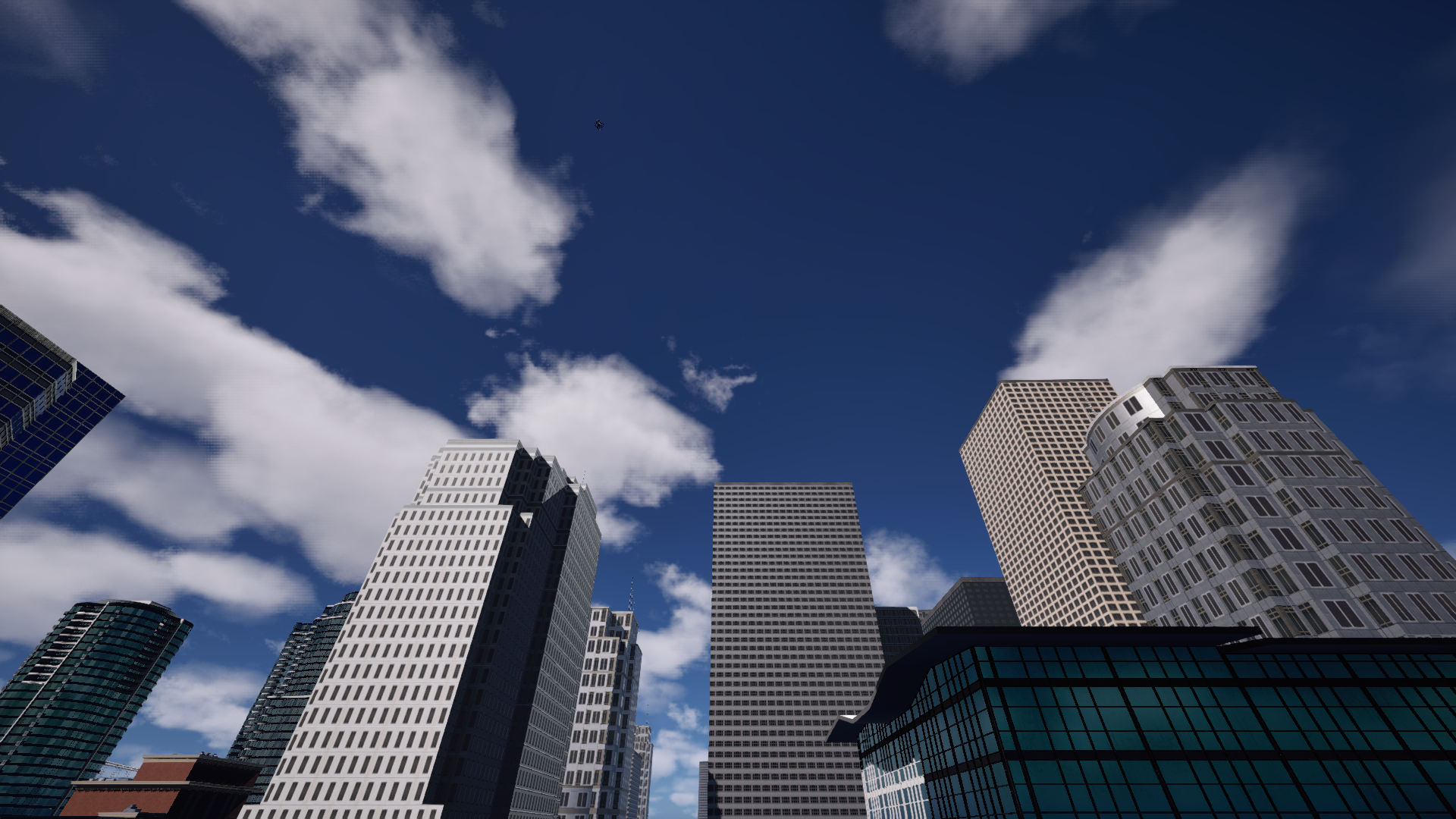}
        \label{fig:scenario2}
    }

    \vspace{0.2em}

    \subfloat[Scene 3]{
        \includegraphics[width=0.48\linewidth]{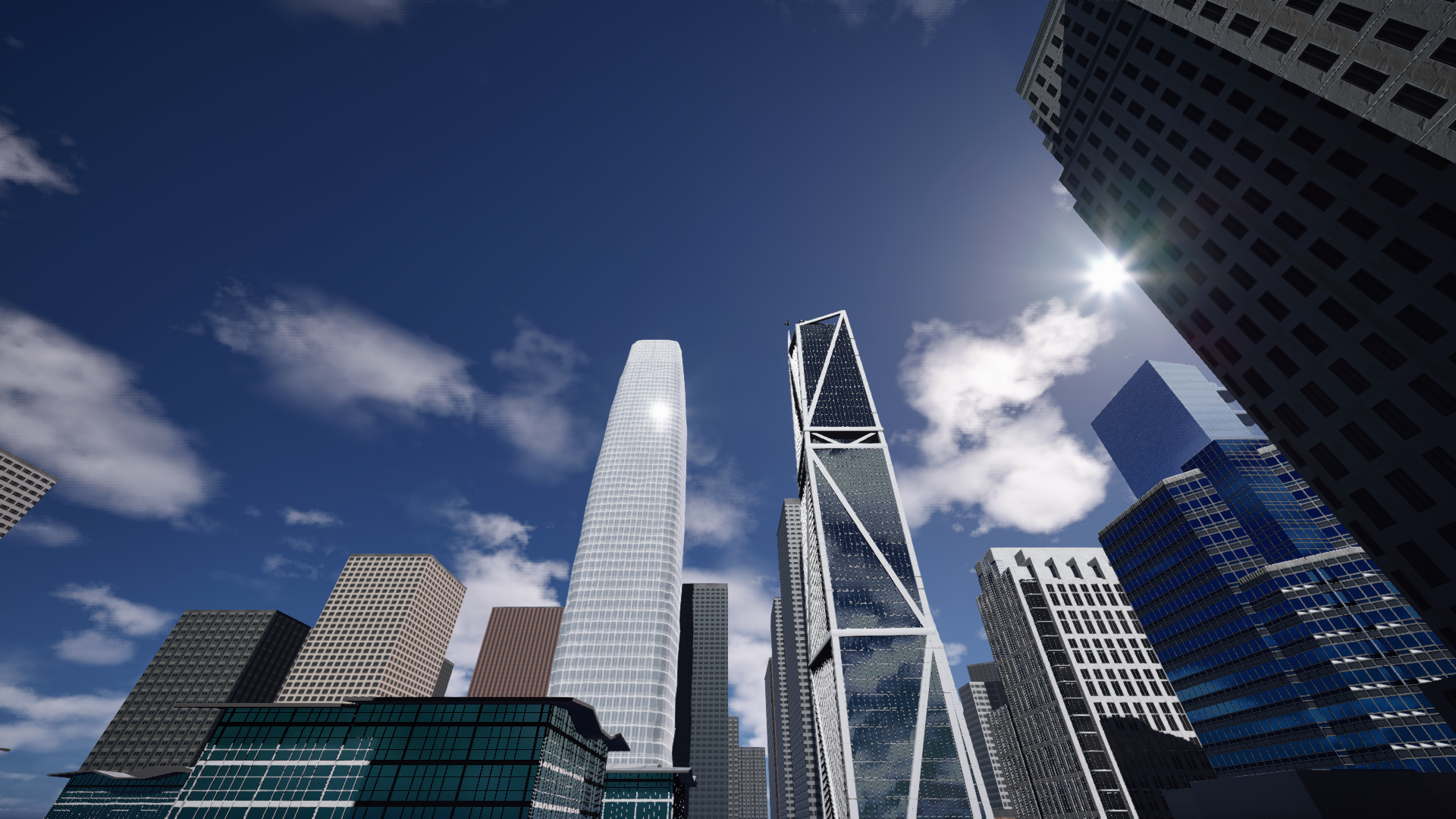}
        \label{fig:scenario3}
    }
    \subfloat[Scene 4]{
        \includegraphics[width=0.48\linewidth]{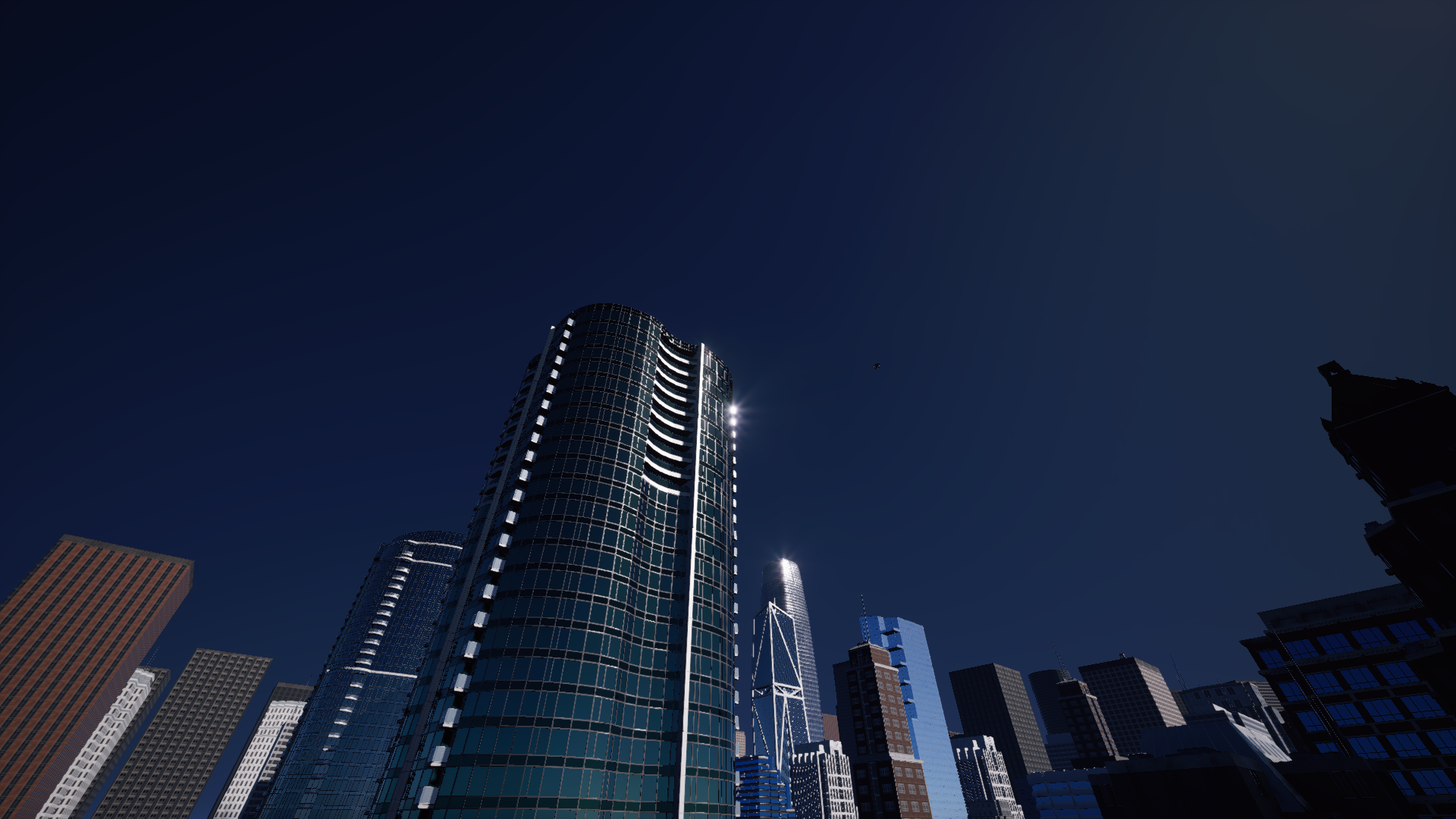}
        \label{fig:scenario4}
    }

    \caption{Examples of the urban scenario subset used in this work. The four scenarios cover different weather, illumination, and UAV trajectory configurations.}
    \label{fig:urban_scenarios}
\end{figure}

The selected subset contains four scenarios with different weather, illumination, and UAV trajectories. Scenarios 1 and 2 are collected under clear daytime and cloudy daytime conditions with trajectory Type A, respectively. Scenario 3 adopts cloudy daytime conditions with trajectory Type B, while Scenario 4 is collected under clear dusk conditions with trajectory Type A. Each scenario includes UAV-camera distances of 70 m, 80 m, 90 m, and 100 m. Representative examples are shown in Fig.~\ref{fig:urban_scenarios}, and the detailed configurations are summarized in Table~\ref{tab:urban_subset}.

\begin{table}[t]
    \centering
    \caption{Urban scenario subset used in the experiments.}
    \label{tab:urban_subset}
    \begin{tabular}{c|c|c|c|c}
        \hline
        \textbf{Scenario} & \textbf{Weather} & \textbf{Time} & \textbf{Trajectory} & \textbf{Distance} \\
        \hline
        1 & Clear & Daytime & Type A & 70--100 m \\
        2 & Cloudy & Daytime & Type A & 70--100 m \\
        3 & Cloudy & Daytime & Type B & 70--100 m \\
        4 & Clear & Dusk & Type A & 70--100 m \\
        \hline
    \end{tabular}
\end{table}

\subsection{Task Definition}

To evaluate the representation capability and transferability of the proposed multimodal foundation model, three downstream tasks are considered in this work: UAV localization, beam prediction, and CSI prediction. These tasks correspond to different levels of low-altitude sensing and wireless intelligence, including spatial perception, communication beam selection, and temporal channel prediction.

\subsubsection{UAV Localization}

UAV localization aims to estimate the 3-D position of the UAV from the available multimodal observations. Given an available modality set \(\mathcal{S}_{ft}\subseteq\mathcal{M}\), the pretrained backbone extracts the fused representation \(\mathbf{F}\), which is then fed into a lightweight regression head to predict the UAV coordinate:
\begin{equation}
    \widehat{\mathbf{y}}
    =
    \mathcal{H}_{loc}
    \left(
    \mathrm{Pool}(\mathbf{F})
    \right),
    \quad
    \widehat{\mathbf{y}}\in\mathbb{R}^{3},
\end{equation}
where \(\mathcal{H}_{loc}(\cdot)\) denotes the localization head. The model is optimized by the regression loss between the predicted and ground-truth UAV positions. This task evaluates whether the fused multimodal representation can preserve spatially discriminative information from visual, geometric, and wireless observations.

\subsubsection{Beam Prediction}

Beam prediction is formulated as a cross-band beam selection task in this work. 
The objective is to infer the optimal 28 GHz mmWave beam index from the aligned 4.9 GHz CSI and multimodal sensing observations. 
This setting exploits the spatial correlation between low-frequency and mmWave channels, where both bands share similar propagation geometry, while avoiding exhaustive beam sweeping at 28 GHz.

Given a predefined beam codebook
\begin{equation}
    \mathcal{B}=\{\mathbf{b}_{1},\mathbf{b}_{2},\ldots,\mathbf{b}_{N_b}\},
\end{equation}
where \(N_b\) is the number of candidate beams, the beam prediction task is formulated as a classification problem:
\begin{equation}
    \widehat{\mathbf{p}}
    =
    \mathcal{H}_{beam}
    \left(
    \mathrm{Pool}(\mathbf{F})
    \right),
    \quad
    \widehat{\mathbf{p}}\in\mathbb{R}^{N_b}.
\end{equation}
The predicted beam is obtained by selecting the class with the maximum probability:
\begin{equation}
    \widehat{b}
    =
    \arg\max_{i}
    \widehat{p}_{i}.
\end{equation}
This task investigates whether the learned multimodal representation can support communication-oriented decision-making by exploiting environmental, geometric, and wireless propagation features.

\subsubsection{CSI Prediction}

CSI prediction aims to forecast the future wireless channel state from the current CSI observation. This task is important for time-varying wireless systems, where accurate channel prediction can reduce channel estimation overhead and support proactive communication optimization. In this work, the input is the CSI matrix at the current time step, and the target is the CSI matrix at the next time step:
\begin{equation}
    \widehat{X}^{c}_{t+1}
    =
    \mathcal{H}_{csi}
    \left(
    \mathrm{Pool}(\mathbf{F}_{t})
    \right),
\end{equation}
where \(\mathbf{F}_{t}\) is the fused feature extracted from the observations at time step \(t\), and \(\widehat{X}^{c}_{t+1}\) is the predicted CSI representation at the next time step.

\subsection{Experimental Design}

The experiments evaluate the proposed multimodal foundation model from four aspects: downstream task performance, missing-modality robustness, pretraining effectiveness, and failure-aware inference. All modality-specific feature extractors are first pretrained on the available training data. The multimodal fusion model is then pretrained on the same data to obtain the foundation backbone. For downstream tasks, the pretrained backbone extracts fixed-size fusion features, followed by a lightweight MLP head for regression or classification.

To reduce experimental complexity, we assume that at most one modality is unavailable or failed at a time. Accordingly, five modality settings are considered, including the full-modality case and four single-missing-modality cases.

\subsubsection{Downstream Task Evaluation}

The pretrained foundation model is evaluated on each scenario under full-sample training and few-shot fine-tuning settings. Its performance is compared with single-modality baselines to verify whether the learned multimodal representation provides stronger transferability and better task performance than individual modality features. All baselines use the same architectures as the corresponding modality feature extractors.

\subsubsection{Missing-Modality Evaluation}

To evaluate robustness under incomplete sensing conditions, the pretrained foundation model is tested with different modality combinations. The five settings include full-modality input and four cases where RGB, depth, LiDAR point cloud, or CSI is missing. This experiment examines whether the backbone can still generate effective fixed-size fusion features when part of the sensing information is unavailable.

\subsubsection{Pretraining Ablation Evaluation}

Ablation experiments are conducted to validate the proposed pretraining strategy. The complete multitask-pretrained model is compared with variants trained using only one pretraining task. A model trained from scratch without pretrained parameters is also included. These comparisons demonstrate the contribution of the pretraining objectives and pretrained backbone to downstream performance.

\subsubsection{Failure-Aware Modality Gating Evaluation}

To further evaluate robustness under modality failure, we test the model with corrupted inputs. Image and depth failures are simulated by occlusion, while LiDAR point cloud and CSI failures are simulated by adding strong noise. The downstream model is trained on normal data and directly tested with one failed modality. We compare the standard M3F-UAV with its Failure-Aware Modality Gating (FAMG) variant to verify whether FAMG can suppress corrupted modality features and prevent them from degrading the fused representation.

\subsection{Implementation Details}

All experiments are conducted on a workstation equipped with an AMD Ryzen 9 9950X3D 16-Core Processor and an NVIDIA GeForce RTX 5090 GPU. The model complexity is reported in Table~\ref{tab:model_complexity}. For all masked feature reconstruction tasks, the masking ratio is set to 50\%. The CSI Transformer and Point Transformer both adopt 4 layers with 4 attention heads, and the feature dimension is set to 512. The MobileNetV2-based visual encoder contains 5 stages, with the final feature dimension of 2560. For the multimodal fusion model, the fusion encoder consists of 4 Transformer layers with 4 attention heads, and the hidden feature dimension is set to 256.

\begin{table}[t]
\centering
\caption{Model complexity of the proposed framework.}
\label{tab:model_complexity}
\small
\setlength{\tabcolsep}{7pt}
\renewcommand{\arraystretch}{1.12}
\begin{tabular}{lccc}
\toprule
\textbf{Module / Model} 
& \textbf{Params} 
& \textbf{FLOPs} 
& \textbf{Inference Latency} \\
& \textbf{(M)} 
& \textbf{(G)} 
& \textbf{(ms/sample)} \\
\midrule
CSI Transformer        & 12.624 & --     & --     \\
Point Transformer      & 13.669 & --     & --     \\
MobileNetV2            & 8.697  & --     & --     \\
M3F-UAV           & 99.138 & 21.938 & 27.667 \\
\bottomrule
\end{tabular}
\end{table}

\subsection{UAV Localization}

\begin{table*}
\centering
\caption{UAV localization performance comparison under single-modality, missing-modality, pretraining-ablation, and fine-tuning settings (NMSE \(\times {10}^2\)). Best and second-best results are highlighted by \textbf{bold} and \protect\uline{underline}.}
\label{tab:localization_nmse}
\small
\setlength{\tabcolsep}{2pt}
\renewcommand{\arraystretch}{1.08}

\begin{tabularx}{\textwidth}{C{0.05\textwidth} *{13}{Y}}
\toprule
\multirow{2}{*}{Scene}
& \multicolumn{4}{c}{Single-modality}
& \multicolumn{4}{c}{Missing-modality}
& \multicolumn{3}{c}{Pretraining-tasks}
& \multicolumn{2}{c}{M3F-UAV} \\
\cmidrule(lr){2-5}
\cmidrule(lr){6-9}
\cmidrule(lr){10-12}
\cmidrule(lr){13-14}
& Image & Depth & LiDAR & CSI
& \makecell{W/o\\Image} 
& \makecell{W/o\\Depth} 
& \makecell{W/o\\LiDAR} 
& \makecell{W/o\\CSI}
& \makecell{W/o\\PT} 
& \makecell{W/o\\Reg} 
& \makecell{W/o\\Rec}
& 10\% & 100\% \\
\midrule

1
& 0.169 & 0.094 & 0.101 & 0.125
& 0.084 & 0.061 & 0.056 & 0.049
& 0.367 & 0.397 & \textbf{0.044}
& 0.060 & \uline{0.048} \\

2
& 0.693 & 0.699 & 1.067 & 0.359
& 0.243 & 0.185 & \uline{0.117} & 0.214
& 0.343 & 0.450 & \textbf{0.094}
& 0.148 & 0.145 \\

3
& 0.566 & 0.523 & 0.809 & 0.304
& 0.222 & 0.304 & 0.288 & 0.287
& 0.441 & 0.516 & \textbf{0.166}
& 0.236 & \uline{0.210} \\

4
& 0.450 & 0.277 & 0.375 & 0.151
& 0.191 & 0.079 & 0.080 & 0.078
& 0.157 & 0.256 & \textbf{0.065}
& 0.133 & \uline{0.076} \\

Avg.
& 0.469 & 0.398 & 0.588 & 0.235
& 0.185 & 0.157 & 0.135 & 0.152
& 0.327 & 0.405 & \textbf{0.092}
& 0.144 & \uline{0.121} \\

\bottomrule
\end{tabularx}
\end{table*}

\subsubsection{Downstream Task Evaluation}

Table~\ref{tab:localization_nmse} compares M3F-UAV with single-modality baselines on the UAV localization task. The single-modality baselines are trained with all available samples, while 100\% and 10\% denote M3F-UAV fine-tuned with all samples and only 10\% samples, respectively.

Compared with single-modality baselines, M3F-UAV achieves the best localization performance in most scenarios, showing that multimodal fusion can effectively exploit complementary visual, geometric, and wireless information. Moreover, M3F-UAV-10\% outperforms all single-modality baselines on average, demonstrating the data efficiency of the pretrained backbone. Scene 1 obtains the lowest errors due to favorable clear daytime conditions, whereas Scenes 2 and 3 show larger errors under cloudy and more challenging sensing environments.

\subsubsection{Missing-Modality Evaluation}

Table~\ref{tab:localization_nmse} also evaluates M3F-UAV under single-missing-modality settings, where ``W/o'' means that the corresponding modality is unavailable during fine-tuning.

Compared with the full-modality setting, most missing-modality variants show only limited performance degradation, and some even achieve better results in certain scenarios. This indicates that the proposed backbone can still generate effective fixed-size fusion representations when one modality is missing.

Among the missing-modality cases, removing LiDAR usually causes relatively small degradation, possibly because UAV-related point-cloud observations are sparse and mainly provide auxiliary geometric cues. In contrast, removing image or depth information often leads to larger errors, suggesting the importance of visual cues for UAV localization.

\subsubsection{Pretraining Ablation Evaluation}

Table~\ref{tab:localization_nmse} further compares different pretraining settings. W/o PT denotes training from scratch without pretrained parameters, while W/o Rec and W/o Reg remove the feature-level reconstruction task and the UAV localization regression task during pretraining, respectively.

W/o Rec achieves the best localization performance, which is reasonable because it is pretrained only with the regression objective and is therefore more aligned with the downstream localization task. Nevertheless, the complete M3F-UAV remains competitive, indicating that feature-level reconstruction does not severely weaken localization performance while improving cross-modal representation learning.

In contrast, W/o PT performs much worse, confirming the importance of pretraining. W/o Reg also suffers from clear degradation, showing that UAV-related localization supervision is necessary for learning task-relevant spatial representations.

\subsubsection{Failure-Aware Modality Gating Evaluation}

\begin{table}[t]
\centering
\caption{Effect of the FAMG module under modality failure conditions on the UAV localization task (NMSE \(\times {10}^2\)). Best results are highlighted by \textbf{bold}.}
\label{tab:FAMG_nmse}
\scriptsize
\setlength{\tabcolsep}{3pt}
\renewcommand{\arraystretch}{1.08}

\begin{tabularx}{\linewidth}{@{}c *{8}{Y}@{}}
\toprule
\multirow{3}{*}{Scene}
& \multicolumn{8}{c}{Failed Modality} \\
\cmidrule(lr){2-9}
& \multicolumn{2}{c}{Image}
& \multicolumn{2}{c}{Depth}
& \multicolumn{2}{c}{LiDAR}
& \multicolumn{2}{c}{CSI} \\
\cmidrule(lr){2-3}
\cmidrule(lr){4-5}
\cmidrule(lr){6-7}
\cmidrule(lr){8-9}
& W/o FAMG & W/ FAMG
& W/o FAMG & W/ FAMG
& W/o FAMG & W/ FAMG
& W/o FAMG & W/ FAMG \\
\midrule

1
& 0.112 & \textbf{0.084}
& 0.079 & \textbf{0.061}
& 0.055 & \textbf{0.046}
& 0.052 & \textbf{0.039} \\

2
& 0.249 & \textbf{0.243}
& 0.304 & \textbf{0.185}
& 0.146 & \textbf{0.117}
& 0.328 & \textbf{0.214} \\

3
& 0.317 & \textbf{0.222}
& 0.421 & \textbf{0.304}
& 0.218 & \textbf{0.188}
& 0.336 & \textbf{0.287} \\

4
& 0.248 & \textbf{0.191}
& 0.108 & \textbf{0.069}
& 0.078 & \textbf{0.070}
& 0.086 & \textbf{0.068} \\

Avg.
& 0.232 & \textbf{0.185}
& 0.228 & \textbf{0.155}
& 0.124 & \textbf{0.105}
& 0.201 & \textbf{0.152} \\

\bottomrule
\end{tabularx}
\end{table}

Table~\ref{tab:FAMG_nmse} compares the localization performance of M3F-UAV with and without the FAMG module under modality failure conditions. Here, the modality listed in the left column denotes the failed modality, ``W/o'' indicates that the corrupted modality feature is still fed into the fusion backbone, and ``W/'' denotes that the failed modality is detected and removed before multimodal fusion.

The results show that FAMG consistently improves localization performance across all failed-modality cases and scenarios. When the image or depth modality fails, FAMG effectively reduces the localization error by suppressing corrupted visual features. Similar improvements are also observed when LiDAR or CSI fails, indicating that unreliable geometric or wireless features may disturb the fused representation if they are directly used for fusion. By removing the failed modality and relying on the remaining reliable modalities, FAMG produces more robust fusion features and improves the accuracy of UAV localization.

\subsection{Beam Prediction}
\begin{table*}
\centering
\captionsetup{labelformat=empty}
\caption{Beam prediction performance comparison under single-modality, missing-modality, pretraining-ablation, and fine-tuning settings (Top-1 and Top-5 Accuracy \%). Best and second-best results are highlighted by \textbf{bold} and \protect\uline{underline}.}
\label{tab:beam_prediction_acc}
\scriptsize
\setlength{\tabcolsep}{0.8pt}
\renewcommand{\arraystretch}{1.08}

\begin{tabularx}{\textwidth}{C{0.032\textwidth} *{26}{Y}}
\toprule
\multirow{3}{*}{Scene}
& \multicolumn{8}{c}{Single-modality}
& \multicolumn{8}{c}{Missing-modality}
& \multicolumn{6}{c}{Pretraining-tasks}
& \multicolumn{4}{c}{M3F-UAV} \\
\cmidrule(lr){2-9}
\cmidrule(lr){10-17}
\cmidrule(lr){18-23}
\cmidrule(lr){24-27}
& \multicolumn{2}{c}{Image}
& \multicolumn{2}{c}{Depth}
& \multicolumn{2}{c}{LiDAR}
& \multicolumn{2}{c}{CSI}
& \multicolumn{2}{c}{\makecell{W/o\\Image}}
& \multicolumn{2}{c}{\makecell{W/o\\Depth}}
& \multicolumn{2}{c}{\makecell{W/o\\LiDAR}}
& \multicolumn{2}{c}{\makecell{W/o\\CSI}}
& \multicolumn{2}{c}{\makecell{W/o\\PT}}
& \multicolumn{2}{c}{\makecell{W/o\\Reg}}
& \multicolumn{2}{c}{\makecell{W/o\\Rec}}
& \multicolumn{2}{c}{10\%}
& \multicolumn{2}{c}{100\%} \\
\cmidrule(lr){2-3}
\cmidrule(lr){4-5}
\cmidrule(lr){6-7}
\cmidrule(lr){8-9}
\cmidrule(lr){10-11}
\cmidrule(lr){12-13}
\cmidrule(lr){14-15}
\cmidrule(lr){16-17}
\cmidrule(lr){18-19}
\cmidrule(lr){20-21}
\cmidrule(lr){22-23}
\cmidrule(lr){24-25}
\cmidrule(lr){26-27}
& T1 & T5
& T1 & T5
& T1 & T5
& T1 & T5
& T1 & T5
& T1 & T5
& T1 & T5
& T1 & T5
& T1 & T5
& T1 & T5
& T1 & T5
& T1 & T5
& T1 & T5 \\
\midrule

1
& 80.89 & 99.13
& 79.63 & 99.31
& 70.59 & 97.59
& \textbf{88.40} & \textbf{99.78}
& 83.22 & 99.06
& 83.65 & 99.08
& 84.26 & 99.08
& 78.93 & 99.23
& 81.92 & 98.61
& 64.78 & 92.34
& 82.40 & 99.26
& 77.16 & 96.48
& \uline{84.41} & \uline{99.28} \\

2
& 83.21 & 99.31
& 69.39 & 88.56
& 45.13 & 73.74
& 93.71 & \textbf{100.0}
& 94.79 & \textbf{100.0}
& 92.38 & \uline{99.98}
& \uline{95.16} & \textbf{100.0}
& 90.39 & 99.80
& 93.79 & \textbf{100.0}
& 88.60 & 99.85
& 94.35 & \textbf{100.0}
& 82.48 & 99.36
& \textbf{95.45} & \textbf{100.0} \\

3
& 85.69 & 99.31
& 66.24 & 92.96
& 43.22 & 71.75
& 89.10 & \uline{99.80}
& 85.44 & 99.36
& 89.39 & 99.59
& \uline{89.55} & 99.73
& 88.15 & 99.70
& 84.88 & 99.48
& 75.55 & 98.34
& 88.09 & 99.48
& 68.06 & 97.02
& \textbf{91.04} & \textbf{99.98} \\

4
& 64.58 & 92.44
& 66.31 & 89.71
& 42.22 & 70.75
& 75.19 & 95.70
& 64.16 & 94.23
& 73.22 & \uline{98.64}
& 75.69 & 98.48
& 53.53 & 95.20
& \uline{76.36} & 98.53
& 47.03 & 83.99
& 55.94 & 87.50
& 51.42 & 89.13
& \textbf{86.04} & \textbf{99.67} \\

Avg.
& 78.59 & 97.53
& 70.39 & 92.63
& 50.29 & 78.46
& \uline{86.62} & 98.82
& 81.90 & 98.16
& 84.66 & \uline{99.32}
& 86.16 & 99.32
& 78.25 & 98.48
& 61.02 & 74.41
& 68.99 & 93.63
& 80.19 & 96.56
& 69.78 & 95.50
& \textbf{89.24} & \textbf{99.73} \\
\bottomrule
\end{tabularx}
\end{table*}

\subsubsection{Downstream Task Evaluation}

Table~\ref{tab:beam_prediction_acc} compares M3F-UAV with single-modality baselines on the beam prediction task using Top-1 and Top-5 accuracy. Overall, M3F-UAV achieves the best performance in most scenarios, especially in Scenes 2--4, where it obtains the highest Top-1 and Top-5 accuracies. This demonstrates that the proposed multimodal foundation model can effectively integrate complementary environmental and wireless features for communication-oriented decision-making.

It is also observed that the CSI baseline is the strongest single-modality method and achieves performance close to M3F-UAV. In Scene 1, CSI even obtains higher Top-1 accuracy than the full multimodal model. This result is reasonable because beam prediction is directly related to wireless channel characteristics, and CSI contains the most task-relevant propagation information for optimal beam selection.

However, M3F-UAV-10\% shows a clear performance drop compared with the fully fine-tuned M3F-UAV. This indicates that beam prediction is more sensitive to downstream labeled data than UAV localization. A possible reason is that beam prediction is a fine-grained classification task, where the decision boundary between adjacent beams can be highly sensitive to small changes in position, orientation, and channel state. 

\subsubsection{Missing-Modality Evaluation}

Table~\ref{tab:beam_prediction_acc} evaluates M3F-UAV under different single-missing-modality settings on the beam prediction task. The full-modality M3F-UAV achieves the best overall performance across all scenarios, demonstrating the effectiveness of fused multimodal features for optimal beam selection. This indicates that visual, geometric, and wireless modalities can provide complementary information for communication-oriented prediction.

When image, depth, or LiDAR is removed, the performance remains close to that of the full-modality model in most cases, showing that the proposed backbone can still extract effective representations under missing-modality conditions. However, removing CSI causes the most obvious performance degradation, especially in Scene 4 where the Top-1 accuracy drops from 77.41\% to 53.53\%. This confirms that CSI is the most critical modality for beam prediction, since it directly reflects wireless propagation and channel characteristics related to the optimal beam.

\subsubsection{Pretraining Ablation Evaluation}

As shown in Table~\ref{tab:beam_prediction_acc}, the complete M3F-UAV achieves the best overall performance on the beam prediction task, reaching the highest average Top-1 and Top-5 accuracies. In contrast, all pretraining ablation variants show performance degradation, indicating that the proposed pretraining strategy is important for learning beam-discriminative multimodal representations. Specifically, W/o PT performs unstably across different scenarios, showing that directly training without pretraining is insufficient for robust beam prediction. W/o Reg also suffers a clear accuracy drop, which suggests that the UAV localization objective provides useful spatial supervision for beam selection. Although W/o Rec performs better than the other incomplete variants, it still remains inferior to the complete M3F-UAV. These results confirm that jointly using reconstruction and UAV localization objectives can better exploit multimodal information and improve beam prediction accuracy.

\subsubsection{Failure-Aware Modality Gating Evaluation}

\begin{table}[t]
\centering
\caption{Effect of the FAMG module under modality failure conditions on the beam prediction task (Top-1 and Top-5 Accuracy \%). Best results are highlighted by \textbf{bold}.}
\label{tab:FAMG_Prediction}
\scriptsize
\setlength{\tabcolsep}{1.2pt}
\renewcommand{\arraystretch}{1.08}

\begin{tabularx}{\linewidth}{@{}c c *{8}{Y}@{}}
\toprule
\multirow{3}{*}{Scene}
& \multirow{3}{*}{Metric}
& \multicolumn{8}{c}{Failed Modality} \\
\cmidrule(lr){3-10}
& & \multicolumn{2}{c}{Image}
& \multicolumn{2}{c}{Depth}
& \multicolumn{2}{c}{LiDAR}
& \multicolumn{2}{c}{CSI} \\
\cmidrule(lr){3-4}
\cmidrule(lr){5-6}
\cmidrule(lr){7-8}
\cmidrule(lr){9-10}
& & w/o FAMG & w/ FAMG
& w/o FAMG & w/ FAMG
& w/o FAMG & w/ FAMG
& w/o FAMG & w/ FAMG \\
\midrule

\multirow{2}{*}{1}
& T1 & 52.06 & \textbf{83.22} & 66.72 & \textbf{83.65} & 76.54 & \textbf{84.26} & 65.47 & \textbf{78.93} \\
& T5 & 90.92 & \textbf{99.06} & 95.45 & \textbf{99.08} & 97.00 & \textbf{99.08} & 93.30 & \textbf{99.23} \\

\midrule
\multirow{2}{*}{2}
& T1 & 77.29 & \textbf{94.79} & 59.82 & \textbf{92.38} & 84.55 & \textbf{95.16} & 15.41 & \textbf{90.39} \\
& T5 & 99.22 & \textbf{100.0} & 95.40 & \textbf{99.98} & 99.60 & \textbf{100.0} & 41.39 & \textbf{99.80} \\

\multirow{2}{*}{3}
& T1 & 28.31 & \textbf{85.44} & 34.20 & \textbf{89.39} & 78.70 & \textbf{89.55} & 51.72 & \textbf{88.15} \\
& T5 & 56.79 & \textbf{99.36} & 81.32 & \textbf{99.59} & 99.06 & \textbf{99.73} & 92.81 & \textbf{99.70} \\

\midrule
\multirow{2}{*}{4}
& T1 & 56.54 & \textbf{64.16} & 47.42 & \textbf{73.22} & 69.69 & \textbf{75.69} & 26.31 & \textbf{55.53} \\
& T5 & 88.19 & \textbf{94.23} & 86.77 & \textbf{98.64} & 97.79 & \textbf{98.48} & 63.47 & \textbf{95.20} \\

\midrule
\multirow{2}{*}{Avg.}
& T1 & 53.55 & \textbf{81.90} & 52.04 & \textbf{84.66} & 77.37 & \textbf{86.16} & 39.73 & \textbf{78.25} \\
& T5 & 83.78 & \textbf{98.16} & 89.73 & \textbf{99.32} & 98.36 & \textbf{99.32} & 72.74 & \textbf{98.48} \\

\bottomrule
\end{tabularx}
\end{table}

Table~\ref{tab:FAMG_Prediction} evaluates the effectiveness of FAMG on the beam prediction task under modality failure conditions. Compared with the setting without FAMG, the model equipped with FAMG consistently achieves much higher Top-1 and Top-5 accuracies across all failed-modality cases. This shows that actively detecting and removing unreliable modality features can effectively mitigate the performance degradation caused by modality failure.

It can also be observed that CSI failure leads to the most severe performance drop when FAMG is not used, especially in Scenes 2 and 4. After introducing FAMG, the performance is significantly recovered, indicating that corrupted CSI features can seriously disturb beam prediction if directly fused. This further confirms the importance of CSI for beam prediction and demonstrates that FAMG improves the robustness of multimodal fusion under practical sensor or signal failure conditions.

\subsection{CSI Prediction}

\begin{table}[t]
\centering
\captionsetup{labelformat=empty}
\caption{Performance comparison of M3F-UAV with the CSI baseline and pretraining ablations on the CSI prediction task (NMSE \(dB\)). Best and second-best results are highlighted by \textbf{bold} and \protect\uline{underline}.}
\label{tab:csi_prediction_nmse}
\footnotesize
\setlength{\tabcolsep}{3pt}
\renewcommand{\arraystretch}{1.08}

\begin{tabular*}{\linewidth}{@{\extracolsep{\fill}}c c c c c c c@{}}
\toprule
\multirow{2}{*}{Scene}
& Single-modality
& \multicolumn{3}{c}{Pretraining-tasks}
& \multicolumn{2}{c}{M3F-UAV} \\
\cmidrule(lr){2-2}
\cmidrule(lr){3-5}
\cmidrule(lr){6-7}
& CSI & W/o PT & W/o Reg & W/o Rec & 10\% & 100\% \\
\midrule
1
& -4.398
& -6.280
& -4.075
& \uline{-7.337}
& -3.523
& \textbf{-7.472} \\

2
& -13.42
& \uline{-16.02}
& -13.92
& -15.41
& -9.442
& \textbf{-17.71} \\

3
& \uline{-11.94}
& -11.55
& -8.587
& -9.873
& -5.772
& \textbf{-13.63} \\

4
& \textbf{-1.986}
& -0.750
& -1.249
& -1.549
& -0.314
& \uline{-1.611} \\

Avg.
& -7.936
& \uline{-8.650}
& -6.958
& -8.542
& -4.763
& \textbf{-10.11} \\
\bottomrule
\end{tabular*}
\end{table}

\subsubsection{Downstream Task Evaluation}

Table~\ref{tab:csi_prediction_nmse} compares M3F-UAV with the CSI single-modality baseline and related ablation variants on the CSI prediction task. Overall, the fully fine-tuned M3F-UAV achieves the best average performance and obtains the best results in most scenarios.

Compared with the CSI baseline, M3F-UAV achieves lower NMSE in Scenes 1--3, indicating that additional environmental modalities provide useful spatial and geometric context for modeling channel evolution. In Scene 4, the CSI baseline performs best, suggesting that CSI alone may be sufficient when the channel dynamics are mainly dominated by wireless observations.

M3F-UAV-10\% shows a clear performance drop compared with the fully fine-tuned model, indicating that CSI prediction is sensitive to downstream labeled data due to its fine-grained temporal and propagation-dependent nature.

\subsubsection{Pretraining Ablation Evaluation}

Among the pretraining ablations, W/o Reg achieves the second-best average performance, while W/o PT shows unstable results across different scenarios. This confirms that pretraining is generally beneficial for CSI prediction, but the contribution of each pretraining objective depends on its consistency with the downstream task. In particular, the complete M3F-UAV achieves the best overall result, indicating that jointly using reconstruction and UAV localization objectives can improve the generality of the learned multimodal representation.

\section{Conclusion}

In this paper, we proposed M3F-UAV, a missing-modality multimodal foundation model for low-altitude wireless sensing. The model integrates RGB images, depth maps, LiDAR point clouds, and CSI matrices through modality-specific pretrained feature extractors and a unified cross-modal fusion backbone, enabling fixed-size feature extraction from different modality combinations. By jointly using feature-level masked reconstruction and UAV localization for pretraining, M3F-UAV learns cross-modal correspondences and UAV-related spatial representations. Experimental results demonstrate its effectiveness in UAV localization and beam prediction, as well as its robustness under missing-modality and modality-failure settings. Future work will extend the framework to more sensing modalities, more challenging multi-modality missing cases, and broader downstream UAV-related wireless tasks.

\bibliographystyle{IEEEtran}
\bibliography{reference}

@article{gupta2016uav_networks,
  author       = {Lav Gupta and
                  Raj Jain and
                  Gabor Vaszkun},
  title        = {Survey of Important Issues in {UAV} Communication Networks},
  journal      = {{IEEE} Commun. Surv. Tutorials},
  volume       = {18},
  number       = {2},
  pages        = {1123--1152},
  year         = {2016}
}

@article{khuwaja2018uav_channel_survey,
  author       = {Aziz Altaf Khuwaja and
                  Yunfei Chen and
                  Nan Zhao and
                  Mohamed{-}Slim Alouini and
                  Paul Dobbins},
  title        = {A Survey of Channel Modeling for {UAV} Communications},
  journal      = {{IEEE} Commun. Surv. Tutorials},
  volume       = {20},
  number       = {4},
  pages        = {2804--2821},
  year         = {2018}
}

@article{yan2023uav_urban_survey,
  author  = {Xiaochen Yan and Tingting Fu and Huaming Lin and Feng Xuan and Yi Huang and Yuchen Cao and Haoji Hu and Peng Liu},
  title   = {{UAV} Detection and Tracking in Urban Environments Using Passive Sensors: A Survey},
  journal = {Applied Sciences},
  volume  = {13},
  number  = {20},
  pages   = {11320},
  year    = {2023}
}

@article{liu2025uav_localization_survey,
  author       = {Haiqiao Liu and
                  Qing Long and
                  Bing Yi and
                  Wen Jiang},
  title        = {A survey of sensors based autonomous unmanned aerial vehicle {UAV} localization techniques},
  journal      = {Complex Intell. Syst.},
  volume       = {11},
  number       = {8},
  pages        = {371},
  year         = {2025}
}

@misc{low_altitude_networks_survey,
      title={Low-Altitude Wireless Networks: A Comprehensive Survey}, 
      author={Jun Wu and Yaoqi Yang and Weijie Yuan and Wenchao Liu and Jiacheng Wang and Tianqi Mao and Lin Zhou and Yuanhao Cui and Fan Liu and Geng Sun and Yiyan Ma and Nan Wu and Dezhi Zheng and Jindan Xu and Nan Ma and Zhiyong Feng and Wei Xu and Dusit Niyato and Chau Yuen and Xiaojun Jing and Zhiguo Shi and Bo Ai and Shi Jin and Dong In Kim and Jiangzhou Wang and Ping Zhang and Hao Yin and Jun Zhang},
      year={2026},
      note={arXiv:2509.11607},
}

@article{lidar_uav_localization,
  author       = {Uthman Olawoye and
                  Jason N. Gross},
  title        = {{UAV} Position Estimation using a {LiDAR}-based 3D Object Detection
                  Method},
  journal      = {CoRR},
  volume       = {abs/2504.07028},
  year         = {2025}
}

@misc{padhy2019uav_localization,
      title={Localization of Unmanned Aerial Vehicles in Corridor Environments using Deep Learning}, 
      author={Ram Prasad Padhy and Shahzad Ahmad and Sachin Verma and Pankaj Kumar Sa and Sambit Bakshi},
      year={2019},
      eprint={arXiv:1903.09021},
}

@inproceedings{charan2022drone_beam_prediction,
  author    = {Charan, Gouranga and Hredzak, Andrew and Stoddard, Christian and Berrey, Benjamin and Seth, Madhav and N{\'u}{\~n}ez, H{\'e}ctor and Alkhateeb, Ahmed},
  title     = {Towards Real-World {6G} Drone Communication: Position and Camera Aided Beam Prediction},
  booktitle = {Proc. {IEEE} Global Commun. Conf. },
  pages     = {2951--2956},
  year      = {2022}
}

@inproceedings{salehi2020camera_beam,
  author       = {Batool Salehi and
                  Mauro Belgiovine and
                  Sara Garcia Sanchez and
                  Jennifer G. Dy and
                  Stratis Ioannidis and
                  Kaushik R. Chowdhury},
  title        = {Machine Learning on Camera Images for Fast mmWave Beamforming},
  booktitle    = {{MASS}},
  pages        = {338--346},
  year         = {2020}
}

@misc{jiang2022lidar_beam,
      title={{LiDAR} Aided Future Beam Prediction in Real-World Millimeter Wave {V2I} Communications}, 
      author={Shuaifeng Jiang and Gouranga Charan and Ahmed Alkhateeb},
      year={2022},
      note={arXiv:2203.05548},
}

@inproceedings{demirhan2022LiDAR_beam,
  author       = {Umut Demirhan and
                  Ahmed Alkhateeb},
  title        = {Radar Aided {6G} Beam Prediction: Deep Learning Algorithms and Real-World Demonstration},
  booktitle    = {Proc. {IEEE} Wireless Commun. Netw. Conf.},
  pages        = {2655--2660},
  year         = {2022},
}

@article{yang_channel_prediction,
  author       = {Heecheol Yang},
  title        = {Deep learning-based channel prediction for {TDD} {MIMO} systems with imperfect channel reciprocity},
  journal      = {{ICT} Express},
  volume       = {11},
  number       = {3},
  pages        = {590--596},
  year         = {2025}
}

@misc{wang2016deepfi,
      title={{CSI}-based Fingerprinting for Indoor Localization: A Deep Learning Approach}, 
      author={Xuyu Wang and Lingjun Gao and Shiwen Mao and Santosh Pandey},
      year={2016},
      eprint={arXiv:1603.07080},
}

@InProceedings{tong2018lstm_channel,
    author="Xiaoyun Tong and Songlin Sun",
    title="Long Short-Term Memory Network for Wireless Channel Prediction",
    booktitle="Proc. Int. Conf. Signal Inf. Process., Netw. Comput. ",
    year="2018",
    pages="19--26",
}

@article{alkhateeb2022deepsense,
  author       = {Ahmed Alkhateeb and
                  Gouranga Charan and
                  Tawfik Osman and
                  Andrew Hredzak and
                  Jo{\~{a}}o Morais and
                  Umut Demirhan and
                  Nikhil Srinivas},
  title        = {DeepSense {6G}: {A} Large-Scale Real-World Multi-Modal Sensing and Communication
                  Dataset},
  journal      = {{IEEE} Commun. Mag.},
  volume       = {61},
  number       = {9},
  pages        = {122--128},
  year         = {2023}
}

@inproceedings{charan2022visionposition,
  author       = {Gouranga Charan and
                  Tawfik Osman and
                  Andrew Hredzak and
                  Ngwe Thawdar and
                  Ahmed Alkhateeb},
  title        = {Vision-Position Multi-Modal Beam Prediction Using Real Millimeter Wave Datasets},
  booktitle    = {Proc. {IEEE} Wireless Commun. Netw. Conf.},
  pages        = {2727--2731},
  year         = {2022},
}

@inproceedings{ahmad2023visiondrone,
  author       = {Iftikhar Ahmad and
                  Ahsan Raza Khan and
                  Rao Naveed Bin Rais and
                  Ahmed Zoha and
                  Muhammad Ali Imran and
                  Sajjad Hussain},
  title        = {Vision-Assisted Beam Prediction for Real World {6G} Drone Communication},
  booktitle    = {Proc. {IEEE} 34th Annu. Int. Symp. Pers., Indoor Mobile Radio Commun.},
  pages        = {1--7},
  year         = {2023},
}

@misc{tian2023multimodal,
      title={Multimodal Transformers for Wireless Communications: A Case Study in Beam Prediction}, 
      author={Yu Tian and Qiyang Zhao and Zine el abidine Kherroubi and Fouzi Boukhalfa and Kebin Wu and Faouzi Bader},
      year={2023},
      eprint={arXiv:2309.11811},
}

@article{yeo2026multimodal,
  author       = {Yerin Yeo and
                  Junghyun Kim and
                  Jihyung Kim and
                  Junhwan Lee},
  title        = {Multi-modal sensing-assisted beam prediction using real-world dataset},
  journal      = {J. Commun. Networks},
  volume       = {27},
  number       = {5},
  pages        = {412--419},
  year         = {2025},
}

@misc{zheng2025m2beamllm,
      title={{M2BeamLLM}: Multimodal Sensing-empowered mmWave Beam Prediction with Large Language Models}, 
      author={Can Zheng and Jiguang He and Chung G. Kang and Guofa Cai and Zitong Yu and Merouane Debbah},
      year={2025},
      note={arXiv:2506.14532},
}

@ARTICLE{xin2024uavchannel,
  author={Xin, Zhichao and Liu, Yu and Xing, Jianping and Huang, Jie and Bian, Ji and Zhang, Yi},
  journal={{IEEE} Internet Things J.}, 
  title={A Novel Multimodal Fusion Sensing-Based Channel Prediction Method for UAV Communications}, 
  year={2025},
  volume={12},
  number={4},
  pages={3948-3960},
}

@inproceedings{brown2020gpt3,
  author       = {Tom B. Brown and
                  Benjamin Mann and
                  Nick Ryder and
                  Melanie Subbiah and
                  Jared Kaplan and
                  Prafulla Dhariwal and
                  Arvind Neelakantan and
                  Pranav Shyam and
                  Girish Sastry and
                  Amanda Askell and
                  Sandhini Agarwal and
                  Ariel Herbert{-}Voss and
                  Gretchen Krueger and
                  Tom Henighan and
                  Rewon Child and
                  Aditya Ramesh and
                  Daniel M. Ziegler and
                  Jeffrey Wu and
                  Clemens Winter and
                  Christopher Hesse and
                  Mark Chen and
                  Eric Sigler and
                  Mateusz Litwin and
                  Scott Gray and
                  Benjamin Chess and
                  Jack Clark and
                  Christopher Berner and
                  Sam McCandlish and
                  Alec Radford and
                  Ilya Sutskever and
                  Dario Amodei},
  title        = {Language Models are Few-Shot Learners},
  booktitle    = {Proc. Adv. Neural Inf. Process. Syst.},
  year         = {2020},
}

@article{chowdhery2022palm,
  author       = {Aakanksha Chowdhery and
                  Sharan Narang and
                  Jacob Devlin and
                  Maarten Bosma and
                  Gaurav Mishra and
                  Adam Roberts and
                  Paul Barham and
                  Hyung Won Chung and
                  Charles Sutton and
                  Sebastian Gehrmann and
                  Parker Schuh and
                  Kensen Shi and
                  Sasha Tsvyashchenko and
                  Joshua Maynez and
                  Abhishek Rao and
                  Parker Barnes and
                  Yi Tay and
                  Noam Shazeer and
                  Vinodkumar Prabhakaran and
                  Emily Reif and
                  Nan Du and
                  Ben Hutchinson and
                  Reiner Pope and
                  James Bradbury and
                  Jacob Austin and
                  Michael Isard and
                  Guy Gur{-}Ari and
                  Pengcheng Yin and
                  Toju Duke and
                  Anselm Levskaya and
                  Sanjay Ghemawat and
                  Sunipa Dev and
                  Henryk Michalewski and
                  Xavier Garcia and
                  Vedant Misra and
                  Kevin Robinson and
                  Liam Fedus and
                  Denny Zhou and
                  Daphne Ippolito and
                  David Luan and
                  Hyeontaek Lim and
                  Barret Zoph and
                  Alexander Spiridonov and
                  Ryan Sepassi and
                  David Dohan and
                  Shivani Agrawal and
                  Mark Omernick and
                  Andrew M. Dai and
                  Thanumalayan Sankaranarayana Pillai and
                  Marie Pellat and
                  Aitor Lewkowycz and
                  Erica Moreira and
                  Rewon Child and
                  Oleksandr Polozov and
                  Katherine Lee and
                  Zongwei Zhou and
                  Xuezhi Wang and
                  Brennan Saeta and
                  Mark Diaz and
                  Orhan Firat and
                  Michele Catasta and
                  Jason Wei and
                  Kathy Meier{-}Hellstern and
                  Douglas Eck and
                  Jeff Dean and
                  Slav Petrov and
                  Noah Fiedel},
  title        = {{PaLM}: Scaling Language Modeling with Pathways},
  journal      = {J. Mach. Learn. Res.},
  volume       = {24},
  pages        = {1--113},
  year         = {2023}
}

@misc{touvron2023llama,
      title={{LLaMA}: Open and Efficient Foundation Language Models}, 
      author={Hugo Touvron and Thibaut Lavril and Gautier Izacard and Xavier Martinet and Marie-Anne Lachaux and Timothée Lacroix and Baptiste Rozière and Naman Goyal and Eric Hambro and Faisal Azhar and Aurelien Rodriguez and Armand Joulin and Edouard Grave and Guillaume Lample},
      year={2023},
      note={arXiv:2302.13971}
}

@ARTICLE{zou2024telecomgpt,
  author = {Hang Zou and Qiyang Zhao and Yu Tian and Lina Bariah and Faouzi Bader and Thierry Lestable and Merouane Debbah},
  journal={{IEEE} Trans. Mach. Learn. Commun. Netw.}, 
  title={{TelecomGPT}: A Framework to Build Telecom-Specific Large Language Models}, 
  year={2025},
  volume={3},
  pages={948-975},
}

@article{liu2024llm4cp,
  author       = {Boxun Liu and
                  Xuanyu Liu and
                  Shijian Gao and
                  Xiang Cheng and
                  Liuqing Yang},
  title        = {{LLM4CP:} Adapting Large Language Models for Channel Prediction},
  journal      = {J. Commun. Inf. Networks},
  volume       = {9},
  number       = {2},
  pages        = {113--125},
  year         = {2024}
}

@ARTICLE{liu2025llm4wm,
  author={Xuanyu Liu and Shijian Gao and Boxun Liu and Xiang Cheng and Liuqing Yang},
  journal={{IEEE} Trans. Mach. Learn. Commun. Netw.}, 
  title={{LLM4WM}: Adapting LLM for Wireless Multi-Tasking}, 
  year={2025},
  volume={3},
  number={},
  pages={835-847},
}

@misc{alikhani2024lwm,
      title={{Large Wireless Model (LWM)}: A Foundation Model for Wireless Channels}, 
      author={Sadjad Alikhani and Gouranga Charan and Ahmed Alkhateeb},
      year={2025},
      note={arXiv:2411.08872},
}

@article{liu2024wifo,
  author       = {Boxun Liu and
                  Shijian Gao and
                  Xuanyu Liu and
                  Xiang Cheng and
                  Liuqing Yang},
  title        = {{WiFo}: wireless foundation model for channel prediction},
  journal      = {Sci. China Inf. Sci.},
  volume       = {68},
  number       = {6},
  year         = {2025}
}

@article{aboulfotouh2025wavesfm,
  author       = {Ahmed AboElfotouh and
                  Elsayed Mohammed and
                  Hatem Abou{-}Zeid},
  title        = {{6G WavesFM}: {A} Foundation Model for Sensing, Communication, and Localization},
  journal      = {{IEEE} Open J. Commun. Soc.},
  volume       = {6},
  pages        = {6792--6807},
  year         = {2025}
}

@inproceedings{chen2024xfi,
  author       = {Xinyan Chen and
                  Jianfei Yang},
  title        = {X-{Fi}: {A} Modality-Invariant Foundation Model for Multimodal Human Sensing},
  booktitle    = {Proc. 13th Int. Conf. Learn. Represent.},
  year         = {2025},
}

@inproceedings{bachmann2022multimae,
  author       = {Roman Bachmann and
                  David Mizrahi and
                  Andrei Atanov and
                  Amir Zamir},
  title        = {{MultiMAE}: Multi-modal Multi-task Masked Autoencoders},
    booktitle = {Proc. Eur. Conf. Comput. Vis.},
  pages        = {348--367},
  year         = {2022},
}

@misc{fu2025physioomni,
      title={Towards Robust Multimodal Physiological Foundation Models: Handling Arbitrary Missing Modalities}, 
      author={Wei-Bang Jiang and Xi Fu and Yi Ding and Cuntai Guan},
      year={2026},
      note={arXiv:2504.19596},
}

@inproceedings{MobileNetV2,
    author       = {Mark Sandler and
                  Andrew G. Howard and
                  Menglong Zhu and
                  Andrey Zhmoginov and
                  Liang{-}Chieh Chen},
    title        = {{MobileNetV2}: Inverted Residuals and Linear Bottlenecks},
    booktitle = {Proc. {IEEE} Conf. Comput. Vis. Pattern Recognit.},
    pages        = {4510--4520},
    year         = {2018},
}

@inproceedings{pointMAE,
  author    = {Pang, Yatian and Wang, Wenxiao and Tay, Francis E. H. and Liu, Wei and Tian, Yonghong and Yuan, Li},
  title     = {Masked Autoencoders for Point Cloud Self-Supervised Learning},
  booktitle = {Proc. Eur. Conf. Comput. Vis.},
  pages     = {604--621},
  year      = {2022}
}

@misc{bian2026airfmddaairinterfacefoundationmodel,
      title={{AirFM-DDA}: Air-Interface Foundation Model in the Delay-Doppler-Angle Domain for {AI}-Native {6G}}, 
      author={Kejia Bian and Meixia Tao and Jianhua Mo and Zhiyong Chen and Leyan Chen},
      year={2026},
      eprint={arXiv:2605.00020},
}

@misc{musefm,
      title={{MUSE-FM}: Multi-task Environment-aware Foundation Model for Wireless Communications}, 
      author={Tianyue Zheng and Jiajia Guo and Linglong Dai and Shi Jin and Jun Zhang},
      year={2026},
      eprint={arXiv:2509.01967},
}

@misc{multimodalmoeisac,
      title={Multimodal Mixture-of-Experts for {ISAC} in Low-Altitude Wireless Networks}, 
      author={Kai Zhang and Wentao Yu and Hengtao He and Shenghui Song and Jun Zhang and Khaled B. Letaief},
      year={2025},
      eprint={arXiv:2512.01750},
}

@misc{jiang2025mimo,
      title={A {MIMO} Wireless Channel Foundation Model via {CIR-CSI} Consistency}, 
      author={Jun Jiang and Wenjun Yu and Yunfan Li and Yuan Gao and Shugong Xu},
      year={2025},
      eprint={arXiv:2502.11965}
}

@article{1,
  author       = {Ahmed AboElfotouh and
                  Elsayed Mohammed and
                  Hatem Abou{-}Zeid},
  title        = {6G WavesFM: {A} Foundation Model for Sensing, Communication, and Localization},
  journal      = {{IEEE} Open J. Commun. Soc.},
  volume       = {6},
  pages        = {6792--6807},
  year         = {2025},
}

\end{document}